\documentclass{aa}  

\usepackage{graphicx}
\usepackage{txfonts}
\begin{document}

   \title{The $\gamma$-ray sky seen at X-ray energies II:}

   \subtitle{the \emph{Swift} hunt of \emph{Fermi} BL Lac objects among unidentified gamma-ray sources}

   \author{E. J. Marchesini
          \inst{1,2,3,4,5}
          \and
          A. Paggi \inst{1}
        \and
        F. Massaro \inst{1}
        \and
        N. Masetti \inst{5,6}
        \and
        R. D'Abrusco \inst{7}
        \and
        I. Andruchow \inst{2,4}
              }

   \institute{Dipartimento di Fisica, Universit\`a degli Studi di Torino, via Pietro Giuria 1, I-10125 Turin, Italy
\and Facultad de Ciencias Astron\'omicas y Geof\'isicas, Universidad Nacional de La Plata, Paseo del Bosque, B1900FWA, La Plata, Argentina
\and INFN -- Istituto Nazionale di Fisica Nucleare, Sezione di Torino, via Pietro Giuria 1, I-10125 Turin, Italy
\and Instituto de Astrof\'isica de La Plata, CONICET--UNLP, CCT La Plata, Paseo del Bosque, B1900FWA, La Plata, Argentina
\and INAF -- Osservatorio di Astrofisica e Scienza dello Spazio, via Gobetti 93/3, I-40129, Bologna, Italy
\and Departamento de Ciencias Fisicas, Universidad Andres Bello, Fernandez Concha 700, Las Condes, Santiago, Chile
\and Center for Astrophysics | Harvard \& Smithsonian , 60 Garden St, Cambridge (MA) 02138, USA
             }

   \date{}

% \abstract{}{}{}{}{} 
% 5 {} token are mandatory
 
  \abstract
   {Nearly 50\% of all sources detected by the \emph{Fermi} Large Area Telescope are classified as blazars or blazar candidates, one of the most elusive classes of active galaxies. Additional blazars can also be hidden within the sample of unidentified or unassociated $\gamma$-ray sources (UGSs) that constitute about one-third of all gamma-ray sources detected to date. We recently confirmed that the large majority of \emph{Fermi} blazars of the BL Lac subclass have an X-ray counterpart.}
   {Using the X-ray properties of a BL Lac training set and combining these with archival multifrequency information, we aim to search for UGSs that could have a BL Lac source within their $\gamma$-ray positional uncertainty regions.}
   {We reduced and analyzed the \emph{Swift} X-ray observations of a selected sample of 327 UGSs. We then compared the X-ray fluxes and hardness ratios of all sources detected in the pointed fields with those of known \emph{Fermi} BL Lacs.}
   {We find at least one X-ray source, lying within the $\gamma$-ray positional uncertainty at 95\%  confidence level, for 223 UGSs and a total of 464 X-ray sources in all fields analyzed. The X-ray properties of a large fraction of them, eventually combined with radio, infrared, and optical information, exhibit BL Lac multi-frequency behavior, thus allowing us to select high-confidence BL Lac candidates; some of them were recently observed during our optical spectroscopic campaign which confirmed their nature.}
   {We find that out of 50 X-ray sources that were confirmed as BL Lacs through optical spectroscopy, 12 do not show canonical mid-infrared or radio BL Lac properties. This indicates that the selection of X-ray BL Lac candidates is a strong method to find new counterparts within \emph{Fermi} UGSs. Finally, we pinpoint a sample of 32 \emph{Swift}/XRT candidate counterparts to \emph{Fermi} UGSs that are most likely BL Lac objects.}


   \maketitle
%
%________________________________________________________________

\section{Introduction} 
\label{sec:intro}

Since its launch in 2008, the Large Area Telescope \citep[LAT,][]{Atwood09} onboard the Fermi satellite, thanks to its superior
angular resolution and larger collecting area with respect to previous 
$\gamma$-ray telescopes, has catalogued $\sim$3000 sources detected to date 
above $\sim$4$\sigma$ significance and listed in the Third Fermi Large Area 
Telescope Source Catalog \citep[3FGL,][]{Acero15}. 

Recently, it has been announced that this number will increase to more than 5000 in the upcoming release of the Fourth Fermi Large Area Telescope Source Catalog \citep[4FGL,][]{4FGL}. Nearly one-third of all sources associated with a low-energy counterpart in the 3FGL belong to the type of active galaxies known as blazars, as was the case with the previous \emph{Fermi} catalogs: 1FGL \citep{Abdo10} and 2FGL \citep{Nolan12}.

Blazars are a peculiar class of active galactic nuclei characterized by emission arising from a relativistic jet oriented at small angles \citep[less than a few degrees; see e.g.,][]{Pushkarev09,Lister13} with respect to the line of sight, which in most cases overwhelms the radiation of its host galaxy \citep{BlandfordRees78}. They are divided into two main classes: flat-spectrum radio quasars with broad emission lines present in their optical spectra, and BL Lac sources, which show featureless optical spectra \citep{Stickel91,Landoni14}.

Two different subclasses for BL Lacs were defined on the basis of the position of the spectral energy distribution (SED) peak for their first component: low-frequency and high-frequency peaked BL Lacs \citep[i.e., HBLs and LBLs;][]{Padovani95}. This criterion is also equivalent to a certain threshold in the ratio between X-ray and radio flux \citep[see also][and references therein]{Maselli10a}.

Another $\sim20$\% of the sources listed in the 3FGL catalog belong to the Blazar Candidate of Uncertain Type (BCU) class, including 
$\gamma$-ray celestial objects having counterparts that show blazar-like characteristics at other wavelengths but still need confirmation of their nature, mainly via optical spectroscopic observations \citep{AlvarezCrespo16,Opt9,DAbrusco19}. Furthermore, an additional $\sim$10\% of the sources listed in 3FGL have Galactic origin, mainly pulsars, pulsar wind nebulae, and supernova remnants \citep[see e.g.,][and references therein]{Massaro15c,Massaro15d}.

The remaining one-third of the 3FGL sources are still unassociated/unidentified. These sources, which also amount to nearly one-fourth in the preliminary release of the 4FGL, are labeled as Unidentified/Unassociated Gamma-ray Sources (UGSs), lacking an assigned low-energy counterpart \citep{Massaro13b,Massaro13e}. Searching for low-energy sources potentially associated with the remaining UGSs is still a challenging task that requires extensive multifrequency follow-up campaigns \citep[e.g.,][]{Acero13,Marchesini16,Lico16} alongside statistical analyses \citep{Ackermann12,Hassan13,DAbrusco14,DAbrusco19}.

Several methods have been developed to search for UGS counterparts. These are based, for example, on optical polarization \citep{Blinov18}, optical spectroscopy \citep{Sandrinelli13,Paiano17b}, $\gamma$-ray spectral and variability characteristics \citep{Ackermann12b}, or on machine learning algorithms \citep{Doert14,Salvetti17}. There are also methods that rely on radio \citep{Healey07,Ghirlanda10,Hovatta12,Nori14,Massaro13e,Nori14} or infrared \citep{DAbrusco12,DAbrusco13} observations. In particular, the latter two methods are based on well-known connections between the blazar emission in these last two bands, and their $\gamma$-ray properties \citep[see, e.g.,][and references therein]{Taylor07,Mahony10,Ackermann11,Massaro16}. 

Nevertheless, optical spectroscopy is still the only method that can determine the real nature of the associated counterpart, and spectroscopic campaigns confirm that most of the UGSs are indeed blazars, in particular of the BL Lac kind \citep{Opt1,Opt3,Opt4,Falomo17,Landoni18,Kaur19,Franceschini19}.

X-ray follow up observations have  also  been widely used as a tool to search for X-ray counterparts that could be blazars for the UGSs   \citep[e.g.,][]{Stephen10,Takahashi12,Maselli13,Paggi13,Masetti13b,Massaro14,Landi15,Marchesini16,Paiano17b}. These X-ray observations were carried out even if a connection between X-ray and $\gamma$-ray emission in blazars was not established as in other energy ranges.

We recently analyzed a uniform sample of \emph{Fermi} BZBs defined by Roma-BZCAT \citep{Massaro15} as BL Lac objects that have been confirmed as such through optical spectroscopy. We proved that, above the $\gamma$-ray flux (i.e., $F_{\gamma}$) threshold of the order of $10^{-11}\,\rm{erg}\,\rm{cm}^{-2}\,\rm{s}^{-1}$, all known $\gamma$-ray BZBs have an X-ray counterpart detected by \emph{Swift}/XRT \citep{Burrows05}, with $\sim$5 ks exposure time and with signal to noise ratios (S/Ns) larger than 3 \citep[][hereinafter paper I]{Marchesini19A}. This X-ray-$\gamma$-ray connection, even if different from those observed at radio and mid-infrared frequencies, supports ongoing and future X-ray follow up observations of UGSs, aiming to find BZB-like counterparts\footnote{https://www.swift.psu.edu/unassociated/} \citep{Stroh13,Falcone14}. It is worth noting that follow-up spectroscopic observations of the X-ray source lying within the positional uncertainty region of each UGS are always necessary to confirm its nature and to potentially obtain its redshift \citep{Massaro15b,Massaro16b,Paiano17a,Paiano19}. 

In this work, we analyze soft X-ray (i.e., in the 0.5-10 keV energy range) observations of Fermi UGSs carried out thanks to the ongoing \emph{Swift}/XRT follow-up campaign \citep{Stroh13,Falcone14}, searching for X-ray sources that show X-ray spectral behaviour similar to that of known and previously analyzed \emph{Fermi} BZBs. We also aim to provide a catalog of X-ray sources that could be targeted by follow up spectroscopic observations to search for UGS counterparts.

This paper is organized as follows. In Section 2 we describe the sample-selection criteria, while in Section 3 we describe the data-reduction procedure. Section 4 is devoted to report our results. Finally, in Section 5 we summarize our main results and report our conclusions.

Throughout the paper, we adopt cgs units and a flat cosmology with $H_0=\,72\,\rm{km}\,\rm{s}^{-1}\,\rm{Mpc}^{-1}$, $\Omega_{\Lambda}=0.74,$ and $\Omega_{m}=0.26$ \citep{Dunkley09}. Spectral indices $\alpha$ were defined so that flux density $\rm{S}_{\nu}\propto\nu^{-\alpha}$, considering $\alpha<0.5$ as flat spectra, especially at radio frequencies around 1.4 GHz. The Wide-field Infrared Survey Explorer \citep[\emph{WISE};][]{Wright10} magnitudes from the All\emph{WISE} catalog, in the 3.4$\mu$m, 4.6$\mu$m, and 12$\mu$m nominal filters, are in the Vega system, and are not corrected for Galactic extinction since this correction is negligible for Galactic latitudes above and below $10^{\circ}$ \citep{DAbrusco13}.

\section{Sample selection} \label{sec:sample}

We selected all \,\emph{Fermi} UGSs listed in the 3FGL \citep{Acero15}, amounting to a total of 1010 sources. We then chose those that had at least one \emph{Swift}/XRT observation performed in photon counting (PC) mode, lying within a circular region of 6 arcmin radius around UGSs positions, for a total of 706 \emph{Fermi} sources. A radius of 6 arcmin was chosen because this is the average semi-major axis of the elliptical positional uncertainty region of UGSs at 95\% level of confidence in the 3FGL, as performed in our previous analysis of Fermi BZBs reported in paper I.

We then selected those with total exposure time in the range between 1 and 10 ks, in agreement with paper I, in which we avoided both under- and over-exposed observations, because the latter are not snapshots. In any case, the aforementioned \emph{Swift} X-ray campaign of UGSs is performed with a nominal 5 ks exposure time \citep{Stroh13}, which is well sampled within our exposure range. Thus, the number of UGSs decreased to 636.

Finally, we discarded all UGSs lying within the Galactic plane ($|b|<10^{\circ}$). This reduces the contamination from Galactic sources. Our final sample consists of 327 \emph{Fermi} UGSs, with an average exposure time of 4.2 ks. The flow chart shown in Figure 1 summarizes all selection steps reported above.

\begin{figure}[ht!]
\centering
\includegraphics[scale=0.75]{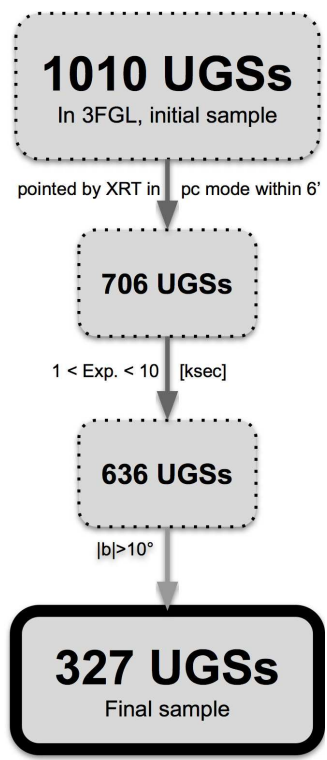}
\caption{Flow chart to highlight all steps followed to build our final UGS sample.}
\end{figure}

\section{\emph{Swift}/XRT data reduction}

\subsection{Data processing}

For all \emph{Swift}/XRT observations analyzed here, we adopted the same data-reduction procedure as in paper I, following \citet{Massaro08b}, \citet{Paggi13} and \citet{Massaro12c}, and references therein. Here we report only basic details.

All \emph{Swift}/XRT observations of our sample of Fermi UGSs were reduced with standard procedures, obtaining clean event files with the use of the \textsc{xrtpipeline} task version 0.13.4, which is part of the \emph{Swift} X-Ray Telescope Data Analysis Software \citep[\textsc{XRTDAS},][]{Capalbi05}. We used the High Energy Astrophysics Science Archive Research Center (HEASARC) calibration database (\textsc{CALDB}) version \textsc{1.0.2}. 

We excluded all time intervals with count rates exceeding 40 counts per second using the \textsc{xselect} task, and all time intervals during which the CCD temperature exceeds $-50^{\circ}$C in regions located at the CCD edge, following recommendations from \citet{DElia13}. Our data-reduction procedure is similar to the one adopted by the \emph{Swift} XRT Point Source catalog \citep[1SXPS,][]{Evans14}, with our results showing differences of only a few percent (see paper I for details).

After extracting all clean event files, we merged exposures of the same source using \textsc{xselect}. The same was done with the corresponding exposure maps with the use of the \textsc{ximage} software.

\begin{figure}[ht!]
\centering
\includegraphics[scale=0.45]{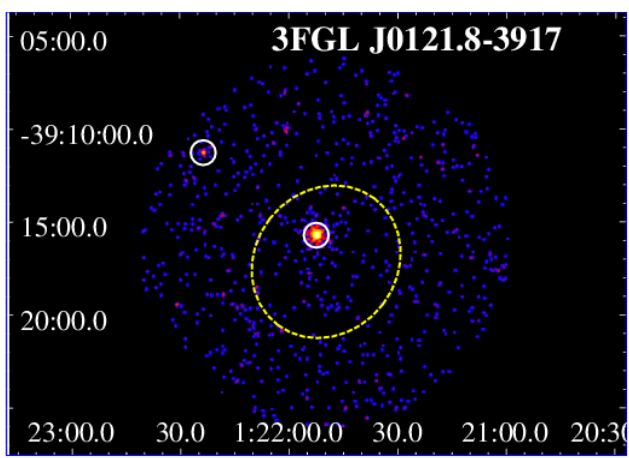}
\caption{Full-band (0.5-10 keV) merged XRT image corresponding to the \,\emph{Fermi} UGS 3FGL\,J0121.8-3917. The yellow dashed line indicates the 3FGL positional uncertainty ellipse at 95\% level of confidence, while the white circles show the XRT sources detected with S/N greater than 3. There is only one XRT counterpart within the positional uncertainty ellipse of 3FGL\,J0121.8-3917. The image was smoothed with a Gaussian kernel radius of 5 pixels.}
\end{figure}

\subsection{Source detection and photometry}

We performed a first detection run over merged event files with the \textsc{det} algorithm in \textsc{ximage}. This way we obtained pixel positions for every detection with S/N larger than 3. 

%The focus of this work is to perform a photometric analysis, thus we did not extract lightcurves or spectra from these sources. Besides, given the selected exposure times we do not expect a sufficient number of counts  (i.e., at least 300 in the 0.5-7 keV energy range) to make a uniform spectral analysis.

We then used the \textsc{sosta} task, available within the \textsc{ximage} package, with the pixel positions obtained from \textsc{det}. Since \textsc{sosta} takes into account the local background and foreground of each X-ray source to evaluate the source photometry, it achieves more precise results than the \textsc{det} algorithm. This way we obtained counts and count rates for each source in the full 0.5-10 keV band, and also in the soft (0.5-2 keV) and hard (2-10 keV) bands. We chose a S/N minimum threshold of 3 to claim a source detection, as in paper I.

We detected 464 X-ray sources within 223 UGS fields. No X-ray sources were found for the remaining 104 UGS, and were consequently excluded from the present analysis. We obtained positions, photon counts, and count rates corrected by the exposure map for every X-ray source. No X-ray sources detected in our sample show pile-up. We then derived the hardness ratio ($\rm{HR}_{\rm{X}}$) for each X-ray source, and full band X-ray fluxes ($\rm{F}_{\rm{X}}$). The hardness ratio was computed as $(\rm{H}-\rm{S})/(\rm{H}+\rm{S})$, where $\rm{H}$ are counts in the \emph{hard} band (2 to 10 keV) and $\rm{S}$ those in the \emph{soft} band (0.5 to 2 keV), respectively. We remark as in our previous work that the exposure map does not change when splitting the data into bands, meaning that using counts or count rates to obtain $\rm{HR}_{\rm{X}}$ is equivalent because the exposure map does not vary between the two energy ranges.

To derive fluxes, we used \textsc{pimms} \citep{Mukai93}, assuming a power-law model with a photon index of 2.0\footnote{The estimate of the X-ray flux only slightly depends on the choice of the photon index for the values typically observed in BZBs as in other radio-loud active galaxies, and it can be considered within the uncertainty of the X-ray flux itself \citep[see e.g.,][and references therein]{Massaro10,Massaro13d,Massaro15d}.} Galactic column density values were taken from the \citet{Kalberla05}. Applying the same criteria reported in paper I, we excluded from our 
sample extended sources and spurious detections due to artifacts and bad pixels, leading to a final sample of 397 XRT sources detected lying within the fields of 223 UGSs. For each X-ray source with equatorial celestial right ascension HH:MM:SS.s and declination $\pm$DD:MM:SS, we adopt a designation of the form SWXRT\,JHHMMSS.s$\pm$DDMMSS, following \citet{Paggi13} as in \citet{DElia13}. In Figure 2 we show the XRT merged image of the \emph{Fermi} UGS 3FGL\,J0121.8-3917, for which we find one XRT counterpart lying within its positional uncertainty ellipse.
In Table 1, we present results obtained from our X-ray data. 

\begin{sidewaystable*}

\caption{First ten rows of the table summarizing the results of our UGS X-ray analysis.}

\begin{center}
\begin{tabular}{|l|c|l|c|c|c|c|c|c|}
\hline
 Name 3FGL &  N & Name XRT & S/N & Soft & Hard & $F_X$  & Ang. sep. & Notes \\

 \, & \,& \, & \,   & counts &  counts & [$\rm{erg}\,\rm{cm}^{-2}\,\rm{s}^{-1}\times10^{-13}$]  & [arcmin] & \\
\hline
\hline

  3FGLJ0006.2+0135 & 1 & SWXRT J000606.0+013121  & 3.9 & 12.5  $\pm$ 4.5  & 0.01 $\pm$0.1 & $1.9\pm0.5$ & 4.7  & N,w,s\\
  3FGLJ0017.1+1445 & 1 & SWXRT J001721.2+145042  & 3.7   & 11.3 $\pm$ 3.9  & 1.1$\pm$1.5  & $1.8\pm0.5$ & 5.7  & w,s\\
  3FGLJ0020.9+0323 & 1 & SWXRT J002116.0+032846  & 3.2 & 10.1 $\pm$ 3.6  & 3.3$\pm$2.1  & $1.4\pm0.5$ & 7.2 & w,s\\
  3FGLJ0031.6+0938 & 1 & SWXRT J003159.7+093617  & 5.0 & 23.9 $\pm$ 5.7  & 3.7$\pm$2.3  & $4.0\pm0.8$ & 5.2 & w,s\\
  3FGLJ0032.3-5522 & 1 & SWXRT J003228.0-551223  & 3.3 & 8.1 $\pm$ 3.4  & 9.5$\pm$3.6  & $0.8\pm0.2$ & 9.8 & w\\
                   & 2 & SWXRT J003149.7-552551  & 3.9 & 12.4 $\pm$ 4.0  & 3.5$\pm$2.5  & $1.0\pm0.3$ & 6.0 & w\\
  3FGLJ0032.5+3912 & 1 & SWXRT J003159.5+391003  & 3.1  & 25.5 $\pm$ 5.1  & 4.1$\pm$2.6  & $1.3\pm0.4$ & 7.4 & w,s \\
                   & 2 & SWXRT J003209.8+392033  & 3.9 & 12.5 $\pm$ 4.1  & 8.1$\pm$3.2  & $2.1\pm0.6$ & 9.2 & w,s \\
  3FGLJ0049.0+4224 & 1 & SWXRT J004859.1+422348  & 7.1 & 48.9 $\pm$ 7.8  & 8.0$\pm$3.4  & $7.0\pm0.9$ & 1.0 & N,w,s\\
  3FGLJ0121.8-3917 & 1 & SWXRT J012152.5-391544  & 20.3 & 306.1 $\pm$ 19.0 & 107.9$\pm$12.0 & $36.0\pm0.3$ & 1.5 & N,w\\
\hline
\end{tabular}
\tablefoot{In column 1 we report the source name as listed in the 3FGL catalog, in column 2 the XRT counterpart identifier indicating also the number of X-ray sources potentially associated with the $\gamma$-ray object, in column 3 the \emph{Swift}/XRT source designation, in column 4 the signal to noise ratio, in columns 5 and 6 the total counts and their uncertainties in the soft (0.5-2 keV) and hard (2-10 keV) X-ray bands, in column 7 we report the full band $\rm{F}_{\rm{X}}$ together with its 1$\sigma$ uncertainty, assuming a power-law model with a photon index of 2.0, and in column 8 the angular separation between the XRT and 3FGL position centroids. Then in column 9 we report notes on the multifrequency archival data found for each source \citep{Opt9}: we mark N those sources having a radio counterpart in the NRAO VLA Sky Survey \citep[NVSS,][]{Condon98}, w for \emph{WISE} All-Sky Survey Catalog \citep{Wright10}, s for Sloan Digital Sky Survey Data Release 9 \citep[SDSS,][]{Ahn12}, and S for Sydney University Molonglo Sky Survey \citep[SUMSS,][]{Mauch03}.}
\end{center}

\end{sidewaystable*}

\section{Results}

\subsection{Crossmatches}

Searching for BZB that could be potentially associated with selected UGSs, we first restricted the sample considering X-ray sources lying within the Fermi positional uncertainty ellipses drawn at 95\% confidence level. This crossmatch was done with an XRT positional uncertainty circle with a radius of 5.6 arcseconds, which is the maximum positional uncertainty reported for 97\% of all 160250 sources in the 1SXPS catalog that lay outside the Galactic plane and were detected with S/N greater than 3 \citep{Evans14}. This resulted in 197 X-ray sources within the positional uncertainty regions of 154 UGSs. 

We found a single X-ray source lying within the positional uncertainty ellipse for 121 UGSs, while for the remaining 33 UGSs there were multiple X-ray counterparts. In particular, 27 out of these 33 UGSs had two X-ray potential counterparts, while 6 UGSs had 3 to 6 X-ray source counterparts within their $\gamma$-ray positional uncertainty.

In the present analysis, UGSs with more than two X-ray sources lying within the Fermi positional uncertainty regions were not investigated, and will be analyzed in a forthcoming paper together with those lying at Galactic latitudes (i.e., $|b|<10^{\circ}$). As an example, in Figure 3 we show the XRT merged image of the \emph{Fermi} UGS 3FGL\,J0809.3-0941, for which we find two X-ray counterparts lying within its positional uncertainty ellipse at 95\% confidence level.

\begin{figure}[ht!]
\centering
\includegraphics[scale=0.45]{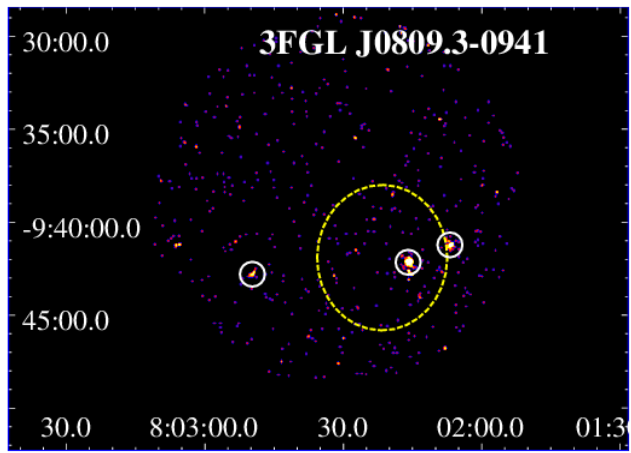}
\caption{Full-band (0.5-10 keV) merged XRT image corresponding to the \,\emph{Fermi} UGS 3FGL\,J0809.3-0941. The yellow dashed line indicates the 3FGL positional uncertainty ellipse at 95\% confidence level, while the white circles show all XRT sources detected with S/N greater than 3. We find two X-ray sources lying within the \emph{Fermi} ellipse, which we label both as XBCs; the remaining one is discarded from this analysis. The image was smoothed with a Gaussian kernel of radius 5 pixels.}
\end{figure}

Our sample lists 148 \emph{Fermi} UGSs having between one and two X-ray sources within their positional uncertainty ellipse at 95\% confidence level, for a total of 175 X-ray sources. This sample is labeled as X-ray blazar candidates (XBCs). 

On the other hand, there are also 69 \emph{Fermi} UGSs with at least one X-ray source detected in the XRT field that does not lie within their positional uncertainty regions. This second sample, labeled as outliers (hereinafter marked as OUTs), lists 105 X-ray sources. We investigate this sample separately from the XBCs, since it is expected that $\sim$5\% of the potential X-ray counterparts do not lie within the 3FGL positional uncertainty ellipse \citep{Acero15,Massaro15c}. The remaining OUTs are probably not associated to the \emph{Fermi} sources, given the fact that we chose to use the positional uncertainty of the 3FGL at 95\% confidence level.

In Figure 4, we show the XRT merged image corresponding to the field of the \emph{Fermi} UGS 3FGL\,J0216.0+0300, in which we find three OUTs. In Figure 5, we show the distribution of the angular separation in arcminutes between the XRT and the 3FGL positions for both subsamples.

It is worth noting that there are X-ray sources in the OUT sample lying at an angular separation comparable to that of XBCs. These could eventually become associated in future releases of the \emph{Fermi} catalogs. Some BZBs could lie outside of the \emph{Fermi} positional uncertainty regions in the current version of the catalog, but a subsequent refined analysis with larger exposure or improved source detection accuracy \citep{Arsioli17} could eventually improve source localization (e.g., due to the improvements in the diffuse model of the gamma-ray background). This could provide future associations, as has occurred in previous versions of the Fermi catalogs \citep{Abdo10,Nolan12,Acero15}.

\begin{figure}[ht!]
\centering
\includegraphics[scale=0.45]{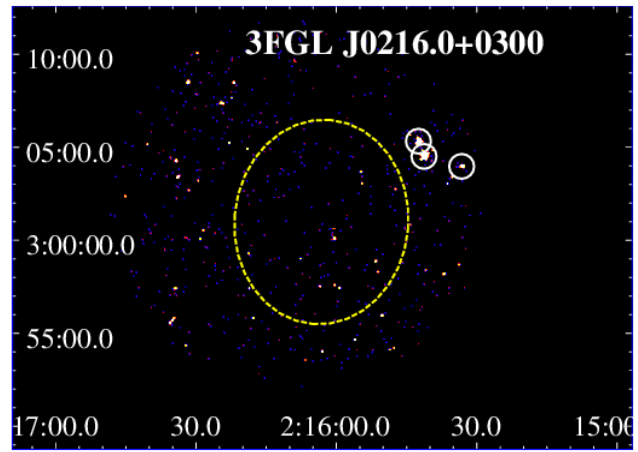}
\caption{Full-band (0.5-10 keV) merged XRT image corresponding to the \,\emph{Fermi} UGS 3FGL\,J0216.0+0300. The yellow dashed line indicates the 3FGL positional uncertainty ellipse at 95\% confidence level, while the white circles show all XRT sources detected with S/N greater than 3. We find three OUTs in this field, since no X-ray source is detected within the \emph{Fermi} ellipse. The image was smoothed with a Gaussian kernel of radius 5 pixels.}
\end{figure}

\begin{figure}[ht!]
\includegraphics[scale=0.45]{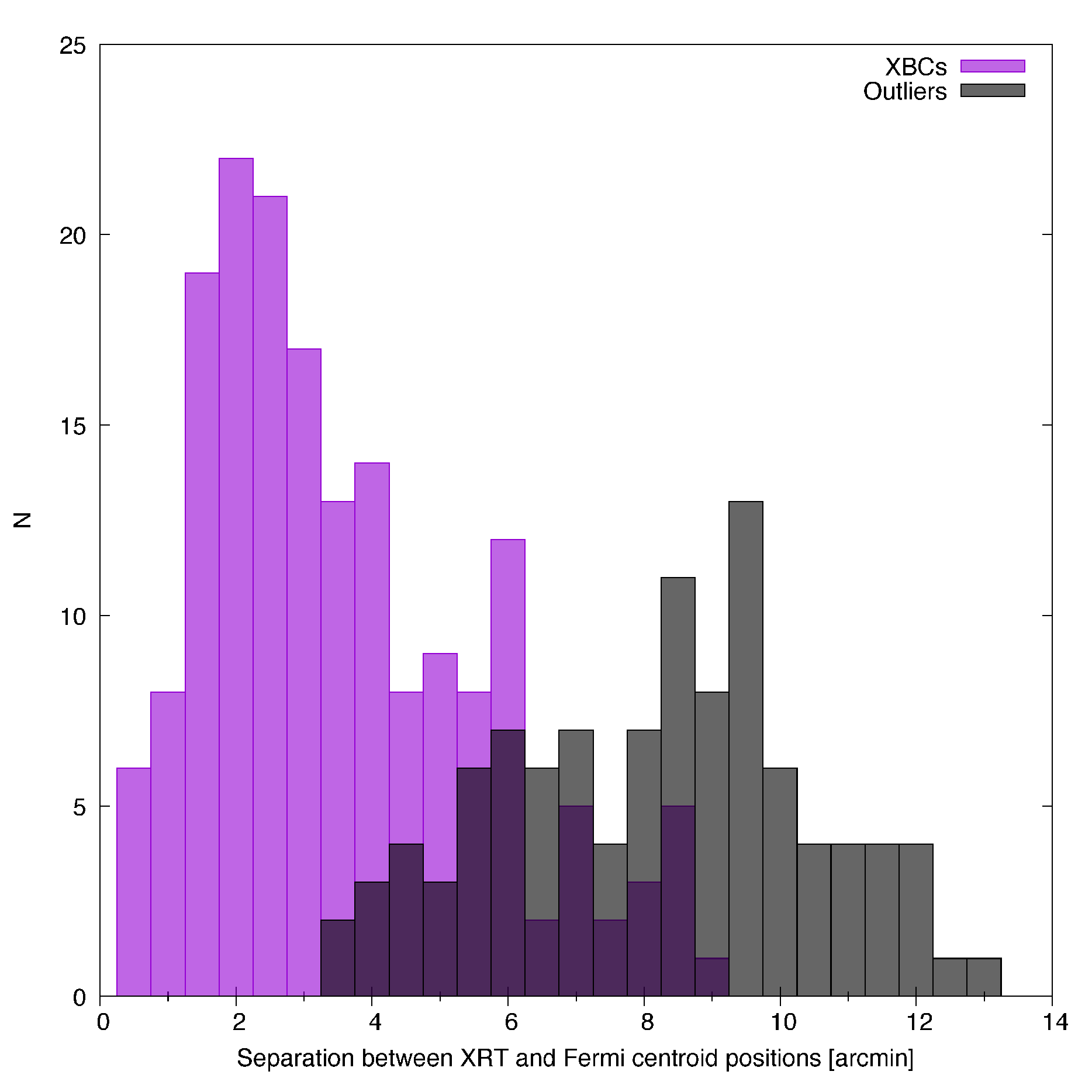}
\caption{Angular separation between \emph{Swift}/XRT and 3FGL positions for XBCs (in purple) and OUTs (in black) in arcminutes. There is an overlap between XBCs and OUTs, implying the latter could be associated in future \emph{Fermi} catalogs.}
\end{figure}

For both subsamples, XBCs and OUTs, we looked for mid-infrared magnitudes in the All\emph{WISE} catalog \citep{Wright10}, and for optical magnitudes in the Sloan Digital Sky Survey \citep[SDSS,]{Ahn12}. We also crossmatched both subsamples with the National Radio Astronomy Observatory Very Large Array Sky Survey catalog \citep[NVSS,][]{Condon98} and the Sydney University Molonglo Sky Survey \citep[SUMSS,][]{Mauch03}.

These crossmatches were done by combining 5.6 arcsec \citep[which is the average positional uncertainty of all sources listed in the 1SXPS catalog][]{Evans14}, with the positional uncertainty of the corresponding catalog. In the case of radio catalogs, we used 10.3 and 7.4 arcseconds for the NVSS and SUMSS catalogs, respectively, as in \citet{DAbrusco14} and \citet{DAbrusco19}. For \emph{WISE} we considered an uncertainty of 3.3 arcseconds, following \citet{DAbrusco13}, while for SDSS we considered an uncertainty of 1.8 arcseconds, following \citet{Massaro14}.

In a total of nine cases between XBCs and OUTs, the crossmatch with \emph{WISE} and SDSS resulted in two counterparts associated to the same XRT source. We chose to keep only the closest one. In total, we obtained 161, 80, and 49 \emph{WISE}, radio, and SDSS unique counterparts for XBCs, and 100, 22, and 34 \emph{WISE}, radio, and SDSS for OUTs, respectively. In particular, 76 and 21 XBCs and OUTs, respectively, have both a radio and a mid-infrared counterpart; and 25 and 5 have  optical, mid-infrared, and radio counterparts, respectively. The results for all crossmatches are shown in Table 2, and in Figure 6 we summarize the selection process for the  two samples in a flow chart.

\begin{table*}
\centering
\caption{Results from the positional crossmatch of XBCs and OUTs with radio, mid-infrared, and optical catalogs.}

\begin{center}

\begin{tabular}{|l|c|c|c|c|c|c|c|}
\hline
                &    &   &    & Radio     & Radio      & Mid-IR     & Radio,   \\
Sample          &   Radio      & Mid-IR        &  Optical         & and & and  & and&  Mid-IR,      \\
                &         &         &           &    Mid-IR        &    Optical        &     Optical         & and Optical      \\
\hline\hline
XBCs            &   80    & 161     &    49     &    76     &       26   &    47        &     25    \\ 
\hline
OUTs        &    22   &   100   &   34      &    21  &     6   &      33          &  5     \\
\hline
\hline\end{tabular}
\end{center}
\end{table*}

%In the end, we found at least one XRT source for 223 of the 327 Fermi UGSs observed by \emph{Swift}/XRT for 1 to 10 ks. Within these 223 UGS fields, we found 397 clean detections with S/N larger than 3. We built a catalog of X-ray Blazar Candidates (XBCs) by taking only those that lied singly or in pairs within the 3FGL position uncertainty ellipse: These amount to 175 out of the 397 XRT sources. Out of the remaining sources, we grouped 105 X-ray sources as ``OUTs'': These correspond to 69 UGSs for which no XRT counterpart was found within its positional uncertainty ellipse. We then looked optical, radio and mid-infrared counterparts for these two samples. 

\begin{figure*}
\centering
\includegraphics[scale=0.70]{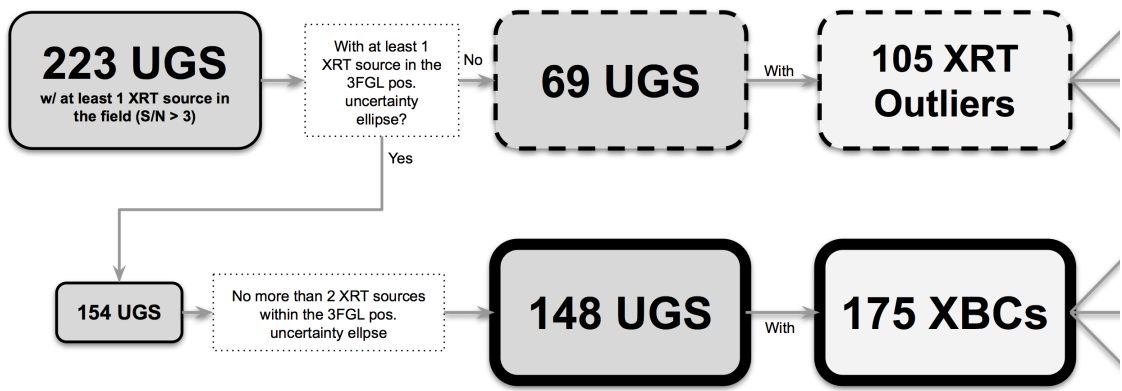}
\caption{Flow chart to highlight all steps followed to build our final XBC and OUT samples.}
\end{figure*}

\subsection{X-ray selection of candidate $\gamma$-ray BL Lacs within the XBCs}

We compared the X-ray properties of XBCs with that of BZBs analyzed in paper I. Although the 3FGL average positional uncertainty of UGSs is of $6.2\pm2.1$ arcminutes, our sample of XBCs show an average angular separation with the 3FGL position of $2.8\pm1.3$ arcminutes, as can be seen in Figure 5. In particular, 93 XBCs lie within less than 3 arcminutes of the 3FGL positional centroid (28\% of all UGSs in our sample). Half of all associated identified BZBs in the 3FGL catalog also lie within 3 arcminutes of the 3FGL positional centroid, suggesting that a fraction of our XBCs could be associated in future releases of the Fermi catalogs as explained in Sec. 4.1.

%This strongly sugggests their possible association in future releases of the Fermi catalogs, given that 285 of all 660 associated/identified BZBs in the 3FGL catalog lie within 3 arcminutes from the 3FGL positional centroid.

%We note that most of our XBC parent sample (55\%) lies within 3.0 arcminutes of the centroid of its corresponding Fermi positional uncertainty ellipse. This fraction increases when taking into account only those XCCs with an associated WISE counterpart (61\%), and when the WISE counterpart lies in the WBS (69\%). In Figure XX we show the angular distance between the position centroid of XRT and that of 3FGL. We show all the 242 XCCs. We also show XCCs with a WISE counterpart, and the ones for which the WISE counterpart lies within the WBS.

In Figure 7 we show $\rm{HR}_{\rm{X}}$ versus $\rm{F}_{\rm{X}}$ for all XBCs (in purple filled circles). We also plot these values for \emph{Fermi} BZBs that we analyzed in paper I as a comparison; these are also reported as background for comparison (i.e., marked as grey filled squares) in all figures shown in the present analysis.

X-ray Blazar Candidates are on average one order of magnitude fainter in X-rays (average flux $\rm{F}_{\rm{X}}=(3.3\pm1.8)\times10^{-13}\,\rm{erg}\,\rm{cm}^{-2}\,\rm{s}^{-1}$) than BZBs $(\rm{F}_{\rm{X}}=(1.2\pm0.9)\times10^{-12}\,\rm{erg}\,\rm{cm}^{-2}\,\rm{s}^{-1})$. This is expected, given the characteristics of each corresponding parent sample: UGSs are in general fainter than identified associated objects in the 3FGL. This is shown and discussed in more detail in Figure 8.

The sample of XBCs shows an average $\rm{F}_{\rm{X}}$ compatible at 1$\sigma$ level with the $\rm{F}_{\rm{X}}$ of what we defined in paper I as `background/foreground objects', that is XRT counterparts not associated with BZBs. These show an average flux of $\rm{F}_{\rm{X}}=(1.0\pm0.4)\times10^{-13}\,\rm{erg}\,\rm{cm}^{-2}\,\rm{s}^{-1}$. 

%fx 

%bzb back    pvalue=3.681277575111492e-61)
%xbc back  pvalue=1.5214867595181968e-13)
%xbc bzb  pvalue=2.5593989402823075e-07), con radio pvalue=2.5593989402823075e-05

%hrx
%bzb back  pvalue=3.647464704634976e-08
%xbc back   pvalue=0.0019104254763296228
%xbc bzb   pvalue=0.5858426178276857)

On the other hand, XBCs show an average $\rm{HR}_{\rm{X}}$ of $-0.56\pm0.18$, while BZBs show an average $\rm{HR}_{\rm{X}}=-0.56\pm0.13$, and background/foreground objects show $\rm{HR}_{\rm{X}}=-0.41\pm0.22$.

We applied a Kolmogorov-Smirnov test to the $\rm{F}_{\rm{X}}$ and $\rm{HR}_{\rm{X}}$ distributions of the XBCs, BZBs, and background/foreground objects in order to test the null hypotesis that the two compared distributions are randomly sampled from a common parent distribution. We found that XBCs and background/foreground objects do not share the same $\rm{F}_{\rm{X}}$ distribution, with a negligible p-chance of $\rm{p}<10^{-13}$, while when comparing XBCs to BZBs we obtain a p-chance of $\rm{p}<10^{-7}$. For background/foreground objects and BZBs, the resulting p-chance is also negligible, being less than $1\times10^{-15}$.

Regarding the distributions of $\rm{HR}_{\rm{X}}$, we find a p-chance of $\rm{p}<0.002$ when comparing XBCs to background/foreground objects, and $\rm{p}<10^{-7}$ when comparing BZBs to background/foreground objects. However, we cannot discard that XBCs and BZBs share a common parent distribution, since when comparing their $\rm{HR}_{\rm{X}}$ distributions we find a p-chance of $\rm{p}\sim0.6$.

The subsample of XBCs with a radio counterpart has an average $\rm{HR}_{\rm{X}}=-0.60\pm0.14$, compatible with that of XBCs. This is an indication that XBCs could belong to the HBL class, since in paper I we found they follow the same pattern, with an average $\rm{HR}_{\rm{X}}=-0.63\pm0.09$. When compared with a Kolmogorov-Smirnov test, we obtain a p-chance of $\rm{p}=0.13$ for these two distributions, which is consistent with them
sharing a common parent distribution.

The discovery of new HBL sources among samples of X-ray and radio-selected objects has  already been indicated in recent studies: the HSP catalogs \citep{Arsioli15,Chang17,Chang19} list HBL sources and candidate sources based on their X-ray-to-radio flux ratios and infrared properties, many of which ($\sim25\%$) are also $\gamma$-ray emitters \citep{Arsioli17,Arsioli18b,Arsioli20}. Indeed, a subsample of our XBCs was listed in the latest version of the HSP catalog, 3HSP. We discuss this sample in more detail in Sect. 4.4.1.

Moreover, for the 27 cases of \emph{Fermi} UGSs with two XBCs lying within the same 3FGL uncertainty ellipse, we plot both sources but we indicate with filled black circles those with the largest angular separation between the X-ray and the UGS $\gamma$-ray position.

For those XBCs with two X-ray sources within the \emph{Fermi} positional uncertainty regions, the most distant X-ray counterpart tends to show $\rm{F}_{\rm{X}}/\rm{HR}_{\rm{X}}$ values that differ from the majority of the XBCs, while the opposite is true for the closest counterparts.

\begin{figure}[ht!]
\includegraphics[scale=0.45,angle=270]{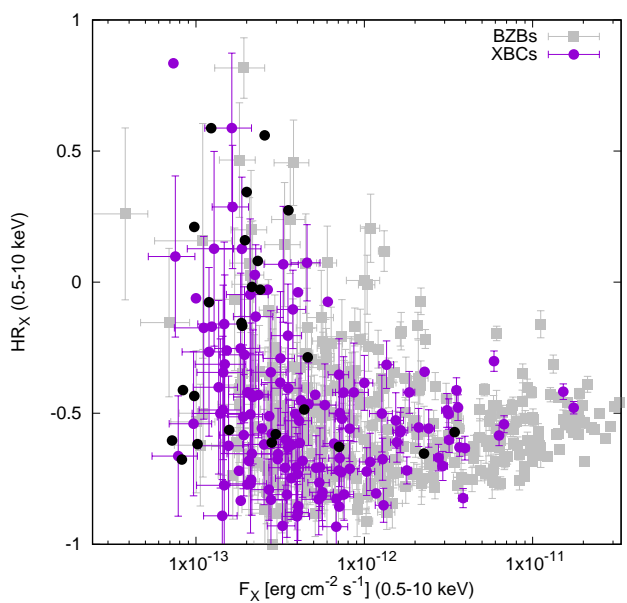}
\caption{$\rm{HR}_{\rm{X}}$ versus $\rm{F}_{\rm{X}}$ in the 0.5-10 keV band for all sources in the XBC sample. XBCs are plotted in empty purple circles, while \emph{Fermi} BZBs are plotted in grey filled squares. When two XBCs lie within the positional uncertainty area of the same UGS, the most distant one is marked with filled black circles.}
\end{figure}

In Figure 8 we show the $\rm{F}_{\gamma}$ in the 100 MeV - 100 GeV band versus $\rm{F}_{\rm{X}}$ for all XBCs. We also compare them to our sample of \emph{Fermi} BZBs. We also report the $\rm{F}_{\gamma}$ thresholds above which 100\%, 98\%, and 96\% of the BZB sample would be detected in X-rays when their exposure times have been scaled down to 5 ks, as shown in paper I. These thresholds are plotted in solid, dashed, and dotted black lines, respectively. They serve as a comparison with BZBs since our XBC sample has an average exposure time of 4.2 ks. 

If the counterpart of an UGS is indeed a BL Lac, above this $\rm{F}_{\gamma}$ threshold we expect to see its X-ray counterpart. Those UGSs emitting above this threshold and lacking X-ray counterparts could probably not be BZBs.

XBCs are fainter in $\gamma$-ray flux than BZBs, as shown in Figure 8. As \emph{Fermi} BZBs analyzed in paper I were the brightest in $\gamma$-rays, while in this work we are sampling the faint tail of the $\gamma$-ray flux distribution, we expect putative BL Lacs in our sample of UGSs to be less bright in X-rays.

In particular, 69\% of the XBCs lie above the 96\% BZB $\rm{F}_{\gamma}$ threshold line. The average $\rm{F}_{\gamma}$ for XBCs is $\rm{F}_{\gamma}=(4.2\pm1.1)\times10^{-12}\,\rm{erg}\,\rm{cm}^{-2}\,\rm{s}^{-1}$, while for the \emph{Fermi} BZBs analyzed in paper I is $\rm{F}_{\gamma}=(8.1\pm3.9)\times10^{-12}\,\rm{erg}\,\rm{cm}^{-2}\,\rm{s}^{-1}$. Since the largest fraction of the XBC sample is fainter in $\gamma$-rays on average than BZBs, this also translates into greater $\gamma$-ray positional uncertainties, which could be the main reason why it is challenging to find lower energy counterparts. There is also no visible trend for those with two X-ray sources within their $\gamma$-ray positional uncertainty region.

%As with BZBs, there is no clear trend, but it is worth noticing that XCCs span on average one order of magnitude in $\gamma$-ray energy flux, in comparison to the sample of BZBs which spans two. In particular, and following our previous analysis, we note that 41\% of the XCCs show $\gamma$-ray fluxes greater than $3.3\times10^{-12}\,\rm{erg}\,\rm{cm}^{-2}\,\rm{s}$, which is the threshold we found for 90\% of our BZBs. Moreover, 20\% of the XCCs show both X-ray and $\gamma$-ray fluxes over the aforementioned thresholds.

\begin{figure}[ht!]
\includegraphics[scale=0.45,angle=270]{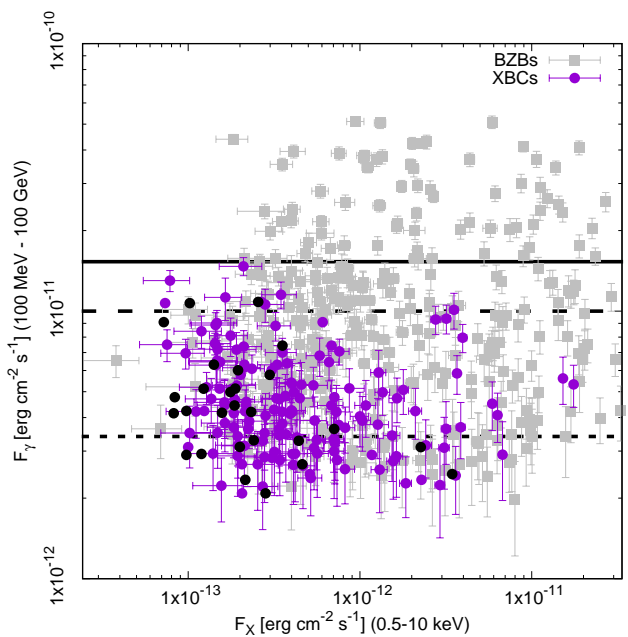}
\caption{ $\gamma$-ray energy flux in the 100 MeV - 100 GeV band vs. $\rm{F}_{\rm{X}}$ from the \,\emph{Fermi} 3FGL catalog for all XBCs in our sample. The symbol coding is the same as in Fig. 7. The $\rm{F}_{\gamma}$ thresholds above which 100\%, 98\%, and 96\% of the BZB sample would be detected in X-rays when their exposure times have been scaled down to 5 ks are plotted in solid, dashed, and dotted black lines, respectively.}

%, in purple empty circles. In grey we plot the comparison sample of BZBs. XBCs that have a radio counterpart are plotted with full purple circles. When two XBCs lie within the positional uncertainty area of the same UGS, the closest one is marked with an empty purple square.}
\end{figure}

In Figure 9 we show the $\gamma$-ray spectral index versus $\rm{HR}_{\rm{X}}$. There is no clear trend for the general XBC sample or for fields with two XBCs.

%The latter is correlated with $\rm{F}_{\rm{X}}$, as we showed in our previous work. In this Figure, the XRT sources with a WISE counterpart outside of the WBS stand out. Nevertheless, most of the sample ($\sim$81\%, including with or without WBS counterparts) follow the correlation we showed exists for Fermi BZBs. Moreover, these sources seem to be closer to the Low-Synchrotron Peak region than to the High-Synchrotron Peak region, as can be seen in the lower panel.

%A comparison between XRT counterparts of UGSs and those of Fermi BZBs. The UGS subsample that has a WISE counterpart is plotted in red, while those whose WISE counterpart lies within the WBS are plotted in blue. Upper panel: X-ray flux in the 0.5-10 keV band versus the W3-W2 color index of the AllWise catalog, where the W3 magnitudes are centered in 12 $\mu$m and W2 in 4.6 $\mu$m. The dashed lines represent the correlations we found in our previous work for a sample of BZBs: for LBLs in orange, for HBLs in teal, and for all BZBs plotted in black. Lower panel: X-ray flux in the 0.5-10 keV band versus the gamma-ray spectral index as reported in the 3FGL catalog, where the x axis is in logarithmic scale. The solid black line indicates the area within the standard deviation of each BZB sample centered in the mean, while the solid grey area indicates the area within the median absolute deviation of each BZB sample centered in the median.

\begin{figure}[ht!]
\includegraphics[scale=0.45,angle=270]{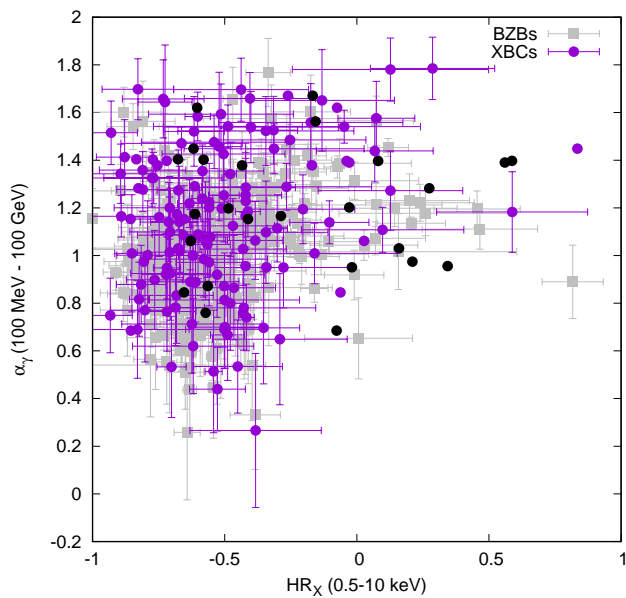}
\caption{ $\gamma$-ray spectral index from the 3FGL vs, $\rm{HR}_{\rm{X}}$ for all XBCs. The symbol coding is the same as in Fig. 7.}
\end{figure}

\subsection{Searching for additional candidate $\gamma$-ray BL Lacs within the OUT sample}

In this section we present the results on \emph{Fermi} UGSs that do not show any X-ray counterpart within their 95\% confidence positional uncertainty ellipse, but show at least one X-ray source out of it.

From our sample of 69 OUTs, 8 ($\sim$12\%) present analysis flags in the 3FGL, all of them indicating uncertainties or artifacts in the $\gamma$-ray analysis due to the adopted $\gamma$-ray diffuse model. However, this is in agreement with XBCs, which also present $\sim$12\% of 3FGL analysis flags. We note that 31 out of 69 OUTs ($\sim$45\%) were listed in the 2FGL catalog, including all 13 ($\sim$19\%) that were listed in the 1FGL catalog.

In Figure 10, we show $\rm{HR}_{\rm{X}}$ versus $\rm{F}_{\rm{X}}$ for all OUTs, and we also show BZBs from paper I. Only a handful of sources lie in the area in which HBLs lie, which corresponds to $\rm{F}_{\rm{X}}\ge10^{-12}$. Their average flux is $\rm{F}_{\rm{X}}=(1.9\pm0.7)\times10^{-13}\,\rm{erg}\,\rm{cm}^{-2}\,\rm{s}^{-1}$, and their average hardness ratio is $\rm{HR}_{\rm{X}}=-0.49\pm0.26$. These are compatible with the XBCs behavior in X-rays, albeit with larger uncertainties. 

%In particular, the average $\rm{HR}_{\rm{X}}$ of OUTs indicates harder spectral shape than XBCs, and is numerically closer to that shown by LBLs in paper I (with an average $\rm{HR}_{\rm{X}}=-0.46\pm0.16$). As before, sources with a radio counterpart (plotted in solid circles) tend to lie in the closer to BZBs in Figure 10.

%Moreover, the catalogs are not \emph{blind}, meaning that a list of sources is provided before analyzing the data, as \emph{seeds} to facilitate the correct source profiling.

\begin{figure}[ht!]
\includegraphics[scale=0.45,angle=270]{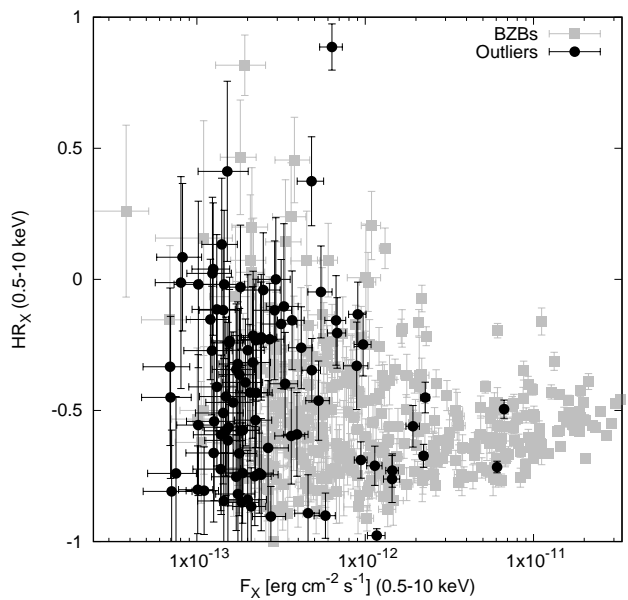}
\caption{$\rm{HR}_{\rm{X}}$ vs. $\rm{F}_{\rm{X}}$ in the 0.5-10 keV band for all the sample of OUTs. OUTs are plotted as empty black circles, while \emph{Fermi} BZBs are plotted as grey filled squares.}
\end{figure}

In Figure 11, we show $\rm{F}_{\gamma}$ in the 100 MeV - 100 GeV band against the full-band $\rm{F}_{\rm{X}}$, again with the thresholds above which 100\%, 98\%, and 96\% of the BZBs should be detected when observed for 5 ks. Of the 105 OUTs, 79 (75\%) display $\rm{F}_{\gamma}$ values above the 96\% threshold. In particular, there are 8 sources for which the $\rm{F}_{\gamma}$ is relatively high, lying above the 100\% threshold line, but their $\rm{F}_{\rm{X}}$ is among the lowest values in the sample. These sources are most probably contaminants of the sample and are likely not associated to UGSs.

\begin{figure}[ht!]
\includegraphics[scale=0.45,angle=270]{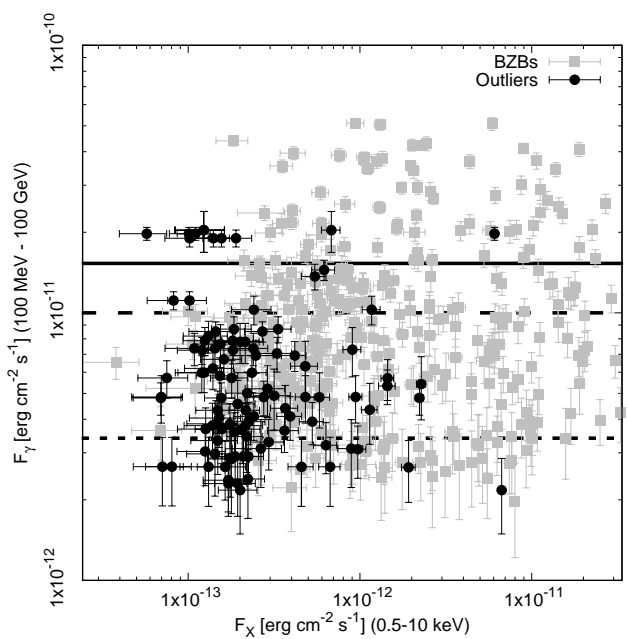}
\caption{$\rm{F}_{\gamma}$ in the 100 MeV - 100 GeV band vs. $\rm{F}_{\rm{X}}$ in the 0.5-10 keV band, from the \,\emph{Fermi} 3FGL catalog, for all OUTs in our sample. The symbol coding is the same as in Fig. 10. The $\rm{F}_{\gamma}$ thresholds above which 100\%, 98\%, and 96\% of the BZB sample would be detected in X-rays when their exposure times have been scaled down to 5 ks, are plotted in solid, dashed and dotted black lines, respectively.}
\end{figure}

\subsection{Multifrequency analysis of candidate $\gamma$-ray BL Lacs}

As previously stated, we crossmatched both the XBC and the OUT subsamples with the All\emph{WISE}, NVSS and SUMSS catalogs: 161 and 100 of XBCs and OUTs, respectively, have a \emph{WISE} mid-infrared counterpart, while 80 and 22 have a radio counterpart. Of the latter, 76 of the XBCs and 21 of the OUTs have both infrared and radio counterparts. Optical, mid-infrared,
and radio  counterparts are present for 25 XBCs and 5 OUTs (see Table 2).

\subsubsection{X-ray Blazar Candidates}

In Figure 12 we show the [3.4]-[4.6] $\mu$m mid-infrared color versus $\rm{F}_{\rm{X}}$. In paper I, we found that BZBs are bluer in mid-infrared when brighter in X-rays. This is also the case for XBCs, although no strict correlation is found in this case. 

More than 80\% of XBCs lie within the BZB area in Figure 12, with only a handful of sources showing [3.4]-[4.6] $\mu$m colors close to zero, which are likely due to contamination from normal elliptical galaxies \citep{Massaro12}.

%In the case of XBCs, we find a mild correlation with a slope of $\sim$-0.5, and a Pearson correlation coefficient of $\rho=-0.40$ for the 161 XBCs with a WISE counterpart. This slope is steeper than that of the \emph{Fermi} BZBs analyzed in paper I, which is $\sim$-0.3. Again, sources with a radio counterpart lie within the BZB area. No clear trend is visible for fields with two XBCs.

\begin{figure}[ht!]
\includegraphics[scale=0.45,angle=270]{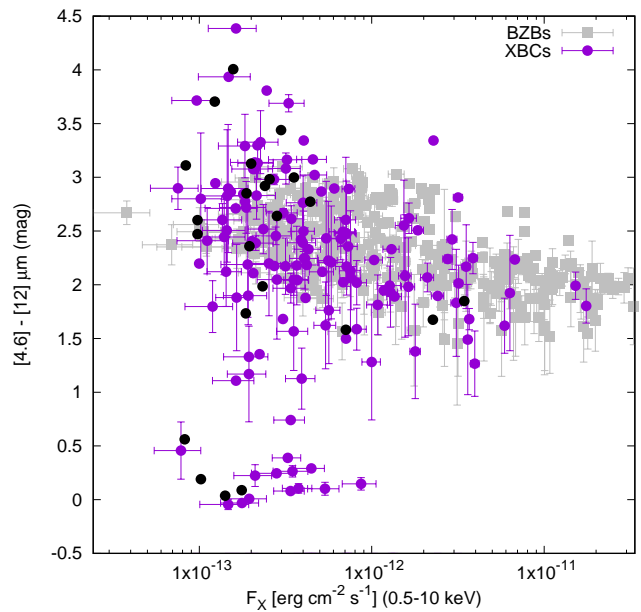}
\caption{The [3.4]-[4.6] $\mu$m mid-infrared color vs. $\rm{F}_{\rm{X}}$ in the 0.5-10 keV band for all  XBCs of the sample with a \emph{WISE} counterpart. The symbol coding is the same as in Fig. 7. Sources with $\mu$m colors close to zero are likely due to contamination from normal elliptical galaxies.}

%The red, blue and grey dashed lines represent the linear fits we applied to the samples of LBLs, HBLs and to all BZBs in paper I. The black dashed line is a fit for the XBC sample, taking out those with mid-infrared colors of elliptical galaxies (i.e. $[3.4]-[4.6]<1$).}

%Those with a radio counterpart are plotted in filled purple circles, while those plotted with empty circles show no radio counterpart. In the cases in which two XBCs lie within the positional uncertainty ellipse of the same \emph{Fermi} UGS, we mark the closest one with empty purple squares. We also plot the sample of \emph{Fermi} BZBs in grey circles.}
\end{figure}

We looked for \emph{WISE} counterparts of XBCs that showed BZB characteristics. \citet{DAbrusco19} built two catalogs of likely mid-infrared counterparts of BZB sources, WIBRALS2, and KDEBLLACS, depending on their infrared and radio properties. We selected \emph{WISE} counterparts of XBCs that were compatible only with the infrared method used to build WIBRALS2 and KDEBLLACS. In particular, WIBRALS2 depends on all four \emph{WISE} bands, including the 22 $\mu$m band, while KDEBLLACS relies only on the 3.4 $\mu$m, 4.6 $\mu$m, and 12 $\mu$m bands. We note that out of the 161 XBCs with a \emph{WISE} counterpart, 74 are detected in all four \emph{WISE} bands while the remaining 87 are not. Out of these, 33 and 49, respectively, are compatible with the infrared color model used to find likely BZBs. Indeed, 23 and 13 are even listed in the WIBRALS2 and KDEBLLACS catalogs, meaning they also comply with the radio selection criteria applied by \citet{DAbrusco19}. Thus, in total, 82 XBCs show BZB-like infrared colors.

%As before, BZBs are plotted in grey circles, and XBCs in empty purple ones. XBCs with a radio counterpart are plotted with filled purple circles.}

We also checked the optical colors of the XBCs. In Figure 13 we show the (u-r) optical color distribution for all 49 XBCs for which we found an SDSS counterpart. There are two sources that show a (u-r) color index greater than 5, not shown in Figure 13. Following \citet{Massaro12}, (u-r) colors lower than 1.4 are a signature of BL Lac sources. Of this
subsample, 34 sources (69\%)   indeed show BL Lac colors, the average being $(u-r)=0.9\pm0.3$.  Of these, 33 have a \emph{WISE} counterpart, 18 (55\%) of which are compatible with BZB-like \emph{WISE} colors. 

Of the 18 XBCs with optical and mid-infrared BZB-like counterparts, 11 also have a radio counterpart. These 11 sources with radio counterparts, (u-r) colors of BL Lacs, and a \emph{WISE} counterpart with BZB-like mid-infrared colors are very likely to be classified as BZBs. 

\begin{figure}[ht!]
\includegraphics[scale=0.45]{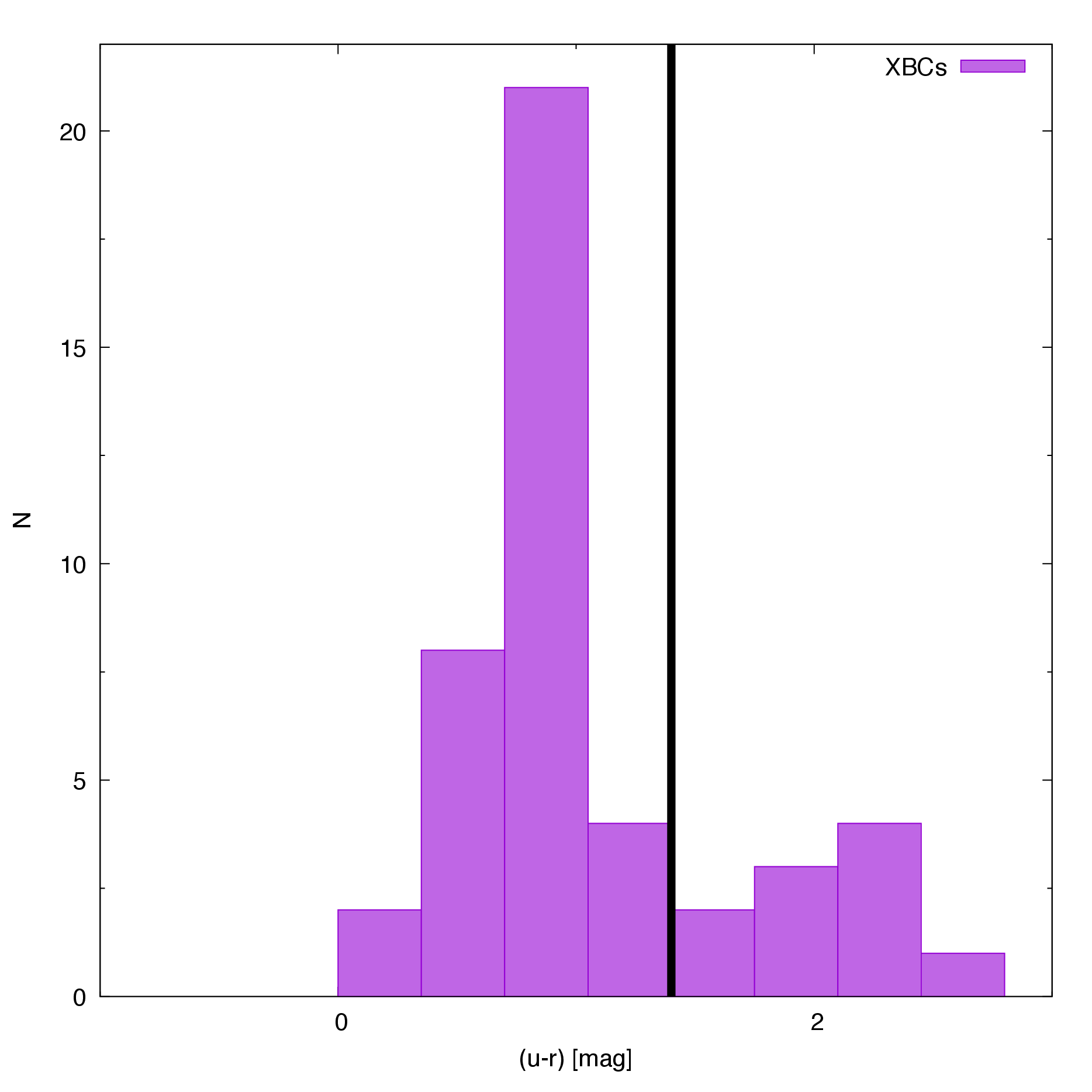}
\caption{Distribution of the (u-r) optical color taken from SDSS for all XBCs with an optical counterpart. The vertical solid line marks the 1.4 limit below which sources are expected to be of the BL Lac type following \citet{Massaro12}.}
\end{figure}

Moreover, 50 of the XBCs were further classified: 3 as quasars (QSO), 2 as BZQs (flat spectrum radio quasars, as defined by Roma-BZCAT), and 45 as  BZBs, all confirmed through optical spectroscopy; the distinction between BZQs and QSOs was made with the use of radio data. These numbers are consistent with our expectations of finding BZBs.

Of these 50 XBCs, 27 were pointed during our ongoing optical spectroscopical campaign aimed at associating counterparts to \emph{Fermi} UGSs \citep{Opt1,Opt2,Opt3,Opt4,Opt5,Opt6,Opt7,Opt8,Opt9}. In these works, we classified sources as BZBs when they showed featureless optical spectra with a dominant blue
continuum \citep[i.e. with emission lines with restframe equivalent widths of less than 5 \AA][]{Laurent98}. We also distinguished BL Lac sources from those with a strong host galaxy contribution, by measuring their relative flux depression towards the CaII line break following \citet{Stocke1991}. Thus, we could ensure the pure BL Lac nature of our sources, all lying within the positional uncertainty ellipse of \emph{Fermi} UGSs.

The remaining 23 XBCs were classified through optical spectroscopy in the literature \citep{Paggi14,Marchesini16,Paiano17b,Marchesi18,Paiano19,Desai19}.

We find that for this subsample of XBCs already associated to BZBs, $(u-r)=1.0\pm0.2$, $\rm{F}_{\rm{X}}=(7.1\pm5.1)\times10^{-13}\,\rm{erg}\,\rm{cm}^{-2}\,\rm{s}^{-1}$, and $\rm{HR}_{\rm{X}}=-0.60\pm0.12$, on average. These values are indeed all consistent with the BZB population, as shown in paper I, and with the whole XBC sample analyzed in this work. This can be seen in Figure 14, where we show again $\rm{HR}_{\rm{X}}$ versus $\rm{F}_{\rm{X}}$ for the XBC sample, but marking the sources that have been classified as BZBs with black circles,  the BZQs with triangles, and the QSOs with diamonds. 

All 45 XBCs classified as BZBs through optical spectroscopy have a mid-infrared counterpart. However,  12 of these BZBs do not display typical mid-infrared BZB colors. Moreover, they are not listed in either the second \emph{WISE} Blazar-like Radio-Loud Sources (WIBRaLS2) catalog or in the KDEBLLACS catalog \citep{DAbrusco19}, which are based on a generalization of the mid-infrared and radio properties of the $\gamma$-ray BL Lac population. We show these 12 BZBs in Table 3. The majority of these sources, 10 out of 12, are listed in the latest catalog of high-synchrotron peaked blazars, 3HSP \citep{Chang19}. This catalog, an updated version of the All\emph{WISE} based 1WHSP \citep{Arsioli15} and 2WHSP \citep{Chang17} catalogs, includes sources with typical HBL X-ray-to-radio flux ratios. Given their selection process, it is probable that these 10 candidates are not only blazars but also of the extreme HBL kind, which would explain their peculiar SED characteristics. This further strengthens the selection of XBCs as a method to find $\gamma$-ray BZBs within \emph{Fermi} UGSs, as $\sim$27\% of the XBCs classified as BZBs do not show canonical mid-infrared or radio properties. 

\begin{table*}
\centering
\caption{\emph{Fermi}, \emph{Swift,} and \emph{WISE} designations for XBCs that were confirmed as BZBs through optical spectroscopy and have a mid-infrared counterpart that is not compatible with the WIBRALS2 or KDEBLLACS color models for typical BZBs.}
\begin{tabular}{|l|l|l|}
\hline
  Name \emph{Fermi} & Name \emph{Swift} & Name \emph{WISE} \\
\hline
  3FGLJ0049.0+4224 & SWXRTJ004859.09+422348.4 & J004859.15+422351.1\\
  3FGLJ0200.3-4108 & SWXRTJ020020.68-410934.9 & J020020.94-410935.7\\
  3FGLJ0704.3-4828 & SWXRTJ070421.64-482645.8 & J070421.81-482647.5\\
  3FGLJ1146.1-0640 & SWXRTJ114600.90-063851.7 & J114600.85-063854.9\\
  3FGLJ1258.4+2123 & SWXRTJ125821.47+212351.7 & J125821.46+212351.1\\
  3FGLJ1411.4-0724 & SWXRTJ141133.30-072254.4 & J141133.31-072253.2\\
  3FGLJ1923.2-7452 & SWXRTJ192241.97-745354.7 & J192243.02-745349.5\\
  3FGLJ2030.5-1439 & SWXRTJ203028.03-143921.2 & J203027.91-143917.1\\
  3FGLJ2034.6-4202 & SWXRTJ203450.87-420037.7 & J203451.08-420038.3\\
  3FGLJ2144.6-5640 & SWXRTJ214429.50-563847.9 & J214429.57-563849.0\\
  3FGLJ2300.0+4053 & SWXRTJ230012.31+405222.6 & J230012.37+405225.1\\
  3FGLJ2321.6-1619 & SWXRTJ232137.01-161925.9 & J232136.98-161928.3\\
  3FGLJ2337.2-8425 & SWXRTJ233624.14-842650.4 & J233627.96-842652.1\\
\hline\end{tabular}
\end{table*}

Moreover, 68\% of XBCs with fluxes $\rm{F}_{\rm{X}}\geq8\times10^{-13}\,\rm{erg}\,\rm{cm}^{-2}\rm{s}^{-1}$ were associated with BZBs. We plot this threshold as a solid black line in Figure 14. We expect that the remaining 32\% of XBCs to be likely BZBs, which could be confirmed through optical follow-up observations.

\begin{figure}[ht!]
\centering
\includegraphics[scale=0.45,angle=270]{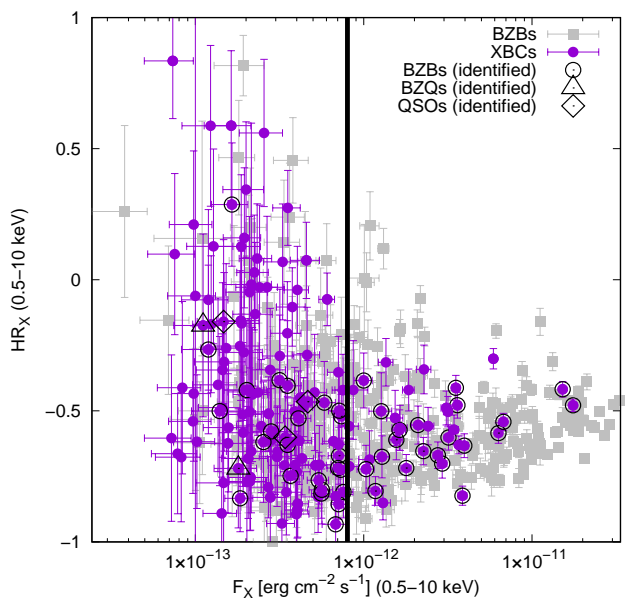}
\caption{$\rm{HR}_{\rm{X}}$ for all the XBC sample vs. $\rm{F}_{\rm{X}}$ in the 0.5-10 keV band. The symbol coding is the same as in Fig. 7. XBCs that have been classified as BZBs are plotted as empty black circles, while BZQs and QSOs as black triangles and diamonds, respectively. The solid black line represents a threshold in $\rm{F}_{\rm{X}}$ above which 68\% of all XBCs have already been identified and only as BZBs.}
\end{figure}

In Figure 15 we show the distribution of $\rm{F}_{\gamma}$ in the 100 MeV - 100 GeV band for the XBCs already classified through optical spectroscopy. We mark again the detection thresholds as in Figure 8. We note that 73\% of this subsample is above the 96\% threshold. This is agreement with the behaviour of the parent XBC sample shown in Figure 8.

Finally, in Figure 16 we show the (u-r) color distribution for the 29 classified XBCs that have a counterpart in SDSS, as in Figure 15. There are six sources above the 1.4 limit, two of which are BZQs and four are BZBs. The remaining sources lie below the 1.4 threshold. Following \citet{Massaro12}, the $\sim$21\% of the sample showing (u-r) color indices above the 1.4 limit can be either high-redshift objects (with $z>0.5$) or galaxy-dominated BZBs, meaning their jets are undergoing a low-activity phase and therefore the host galaxy becomes the dominant feature in its optical spectrum.

\begin{figure}[ht!]
\centering
\includegraphics[scale=0.45]{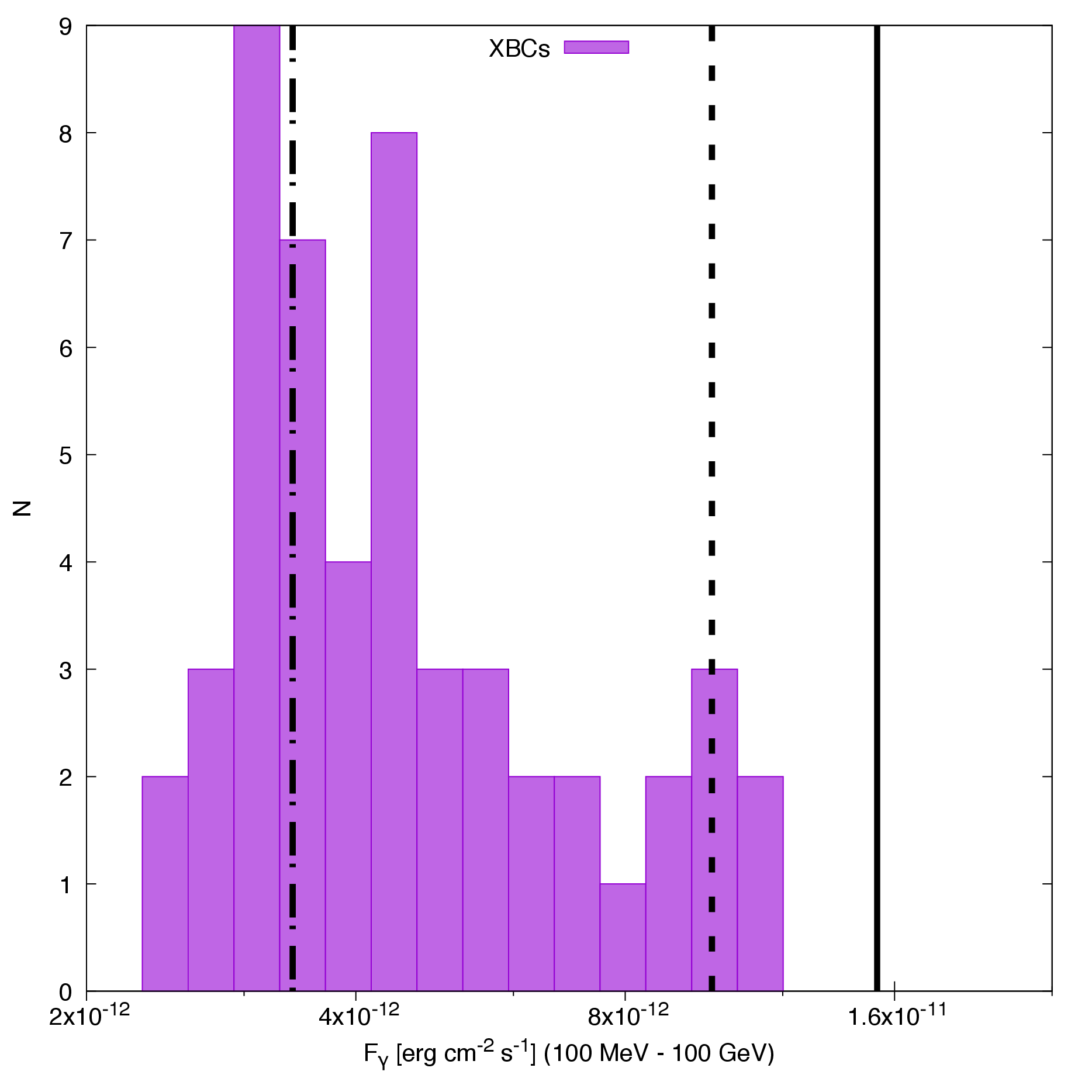}
\caption{Distribution of $\rm{F}_{\gamma}$ in the 100 MeV - 100 GeV band for the XBCs already classified through optical spectroscopy. The thresholds above which 100\%, 98\%, and 96\% of the BZBs are expected to be detected in X-ray when observing for 5 ks or more are plotted as solid, dashed, and dot-dashed black lines, respectively.}
\end{figure}

\begin{figure}[ht!]
\centering
\includegraphics[scale=0.45]{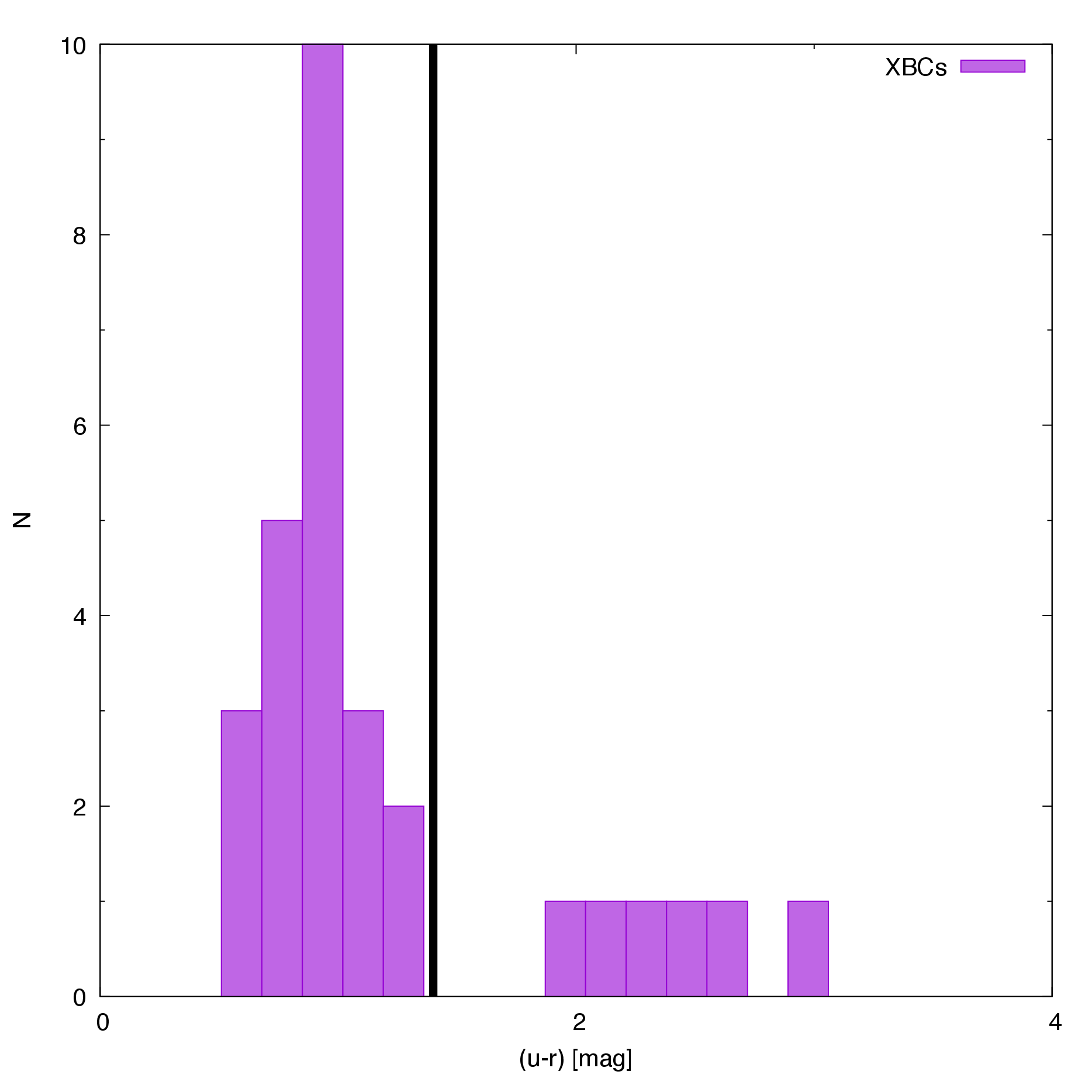}
\caption{Distribution of the (u-r) optical color taken from SDSS for all XBCs with an optical counterpart associated through optical spectroscopy and classified as either BZBs, BZQs, or QSOs. The vertical solid line marks the limit below which sources are expected to be of the BL Lac type following \citet{Massaro12}.}
\end{figure}

The fact that XBCs  closely follow the BZB trends in mid-infrared colors, X-ray hardness ratio values, and optical color indices indicates that they share a similar spectral shape in all these bands with BZBs. The main difference between XBCs and BZBs lies in their fluxes,  XBCs being  fainter than BZBs in both $\gamma$-rays and X-rays.

We selected a subsample of XBCs that are likely BZB sources with a high confidence degree. We selected all XBCs that satisfied {at least} two of the following criteria:

\begin{itemize}
    \item XBCs that have a radio counterpart,
    \item XBCs that have an optical counterpart with color index (u-r)$<$1.4,
    \item XBCs that have a mid-infrared counterpart which is compatible with either the WIBRALS2 or the KDEBLLACS mid-infrared color model for BZBs.
\end{itemize}

Thus, we obtained a subsample of 54 XBCs. Finally, we discarded 35 sources that had  already been associated in the literature or during our follow-up campaigns. The remaining 19 form part of our catalog of likely BZBs that will be part of our future optical spectroscopic follow-up campaigns. 

We highlight the fact that with this selection process, all XBCs correspond to a single UGS field. This means that, in cases in which we found two XBCs within the same UGS uncertainty ellipse, at least one of them did not satisfy two of the three criteria stated above. We list the whole sample of 30 X-ray sources that are candidate $\gamma$-ray BL Lacs in Table 4. In column 1 we report the 3FGL name, in column 2 the \emph{Swift}/XRT source designation, in column 3 the angular separation between the 3FGL and XRT positions, and in column 4 the class we assigned to it. This table includes sources selected from the OUT sample (see Sec. 4.4.2). We note that nine of the sources included in Table 4 also belong to the 3HSP catalog \citep{Chang19}.

\subsubsection{\emph{Outliers}}

The sample of OUTs show greater variation in their properties. In Figure 17, we show the [3.4]-[4.6] $\mu$m mid-infrared color versus $\rm{F}_{\rm{X}}$, again comparing with BZBs and their trends as explained in Figure 12. 

There are 18 sources that show  $[3.4]-[4.6]\,\mu\rm{m}\approx0$, which is the typical color of elliptical galaxies, or of BL Lacs for which the contribution of the host galaxy is dominant. The remaining sources show values similar to BZBs, with all but nine of the OUTs lying towards the fainter side in X-rays, and being redder in mid-infrared ($\rm{F}_{\rm{X}}<1.0\times10^{-12}$, $[3.4]-[4.6]<1.5$). Again, sources with a radio counterpart tend to group together in the area covered by BZBs. 

\begin{figure}[ht!]
\includegraphics[scale=0.45,angle=270]{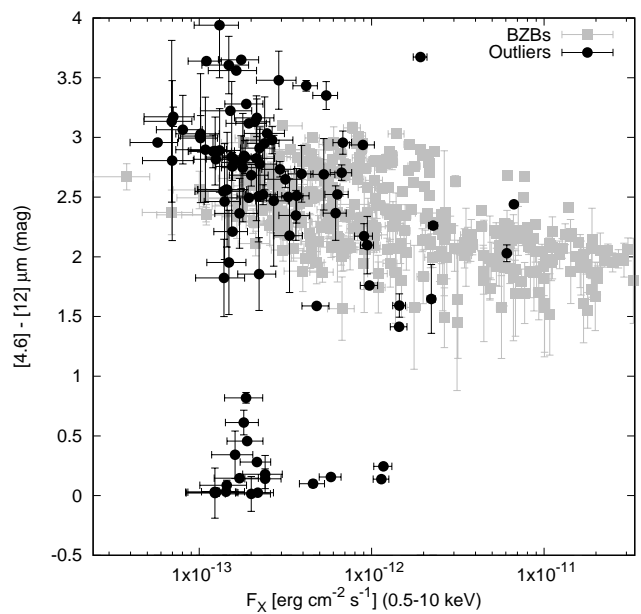}
\caption{The [3.4]-[4.6] $\mu$m mid-infrared color $\rm{F}_{\rm{X}}$ in the 0.5-10 keV band, for all the sample of OUTs with a \emph{WISE} counterpart. The symbol coding is the same as in Fig. 10.}

%Those with a radio counterpart are plotted in filled black circles, while those plotted with empty circles show no radio counterpart. We also plot the sample of \emph{Fermi} BZBs of paper I in grey circles. }
\end{figure}

When checking their mid-infrared colors, 23 OUTs are compatible with the WIBRALS2 color model, while another 13 sources with no detection at 22 $\mu$m are compatible with the KDEBLLACS infrared color model \citep{DAbrusco19}. This means that a total of 36 OUTs display infrared colors of BZBs, with 8 of these having also a radio counterpart; these latter are therefore listed in both the WIBRALS2 and KDEBLLACS catalogs of infrared BZB-like sources.

The majority of the OUT sample, that is, 64 sources, are not compatible with either infrared color model. However, we stress that there is still a non-negligible (at least 30\%) number of potential BZBs within the OUTs sample.

%As before, BZBs are plotted in grey circles, and OUTs in empty black ones. OUTs with a radio counterpart are plotted with filled black circles.}

Regarding the optical colors of OUTs, in Figure 18 we show their (u-r) distribution. There are 34 OUTs with a SDSS counterpart; 14 of these show $(u-r)<1.4$ mag, which is typical of BZBs. All 34 show large scatter, their average color index being $(u-r)=1.7\pm1.2$ mag.

\begin{figure}[ht!]
\centering
\includegraphics[scale=0.45]{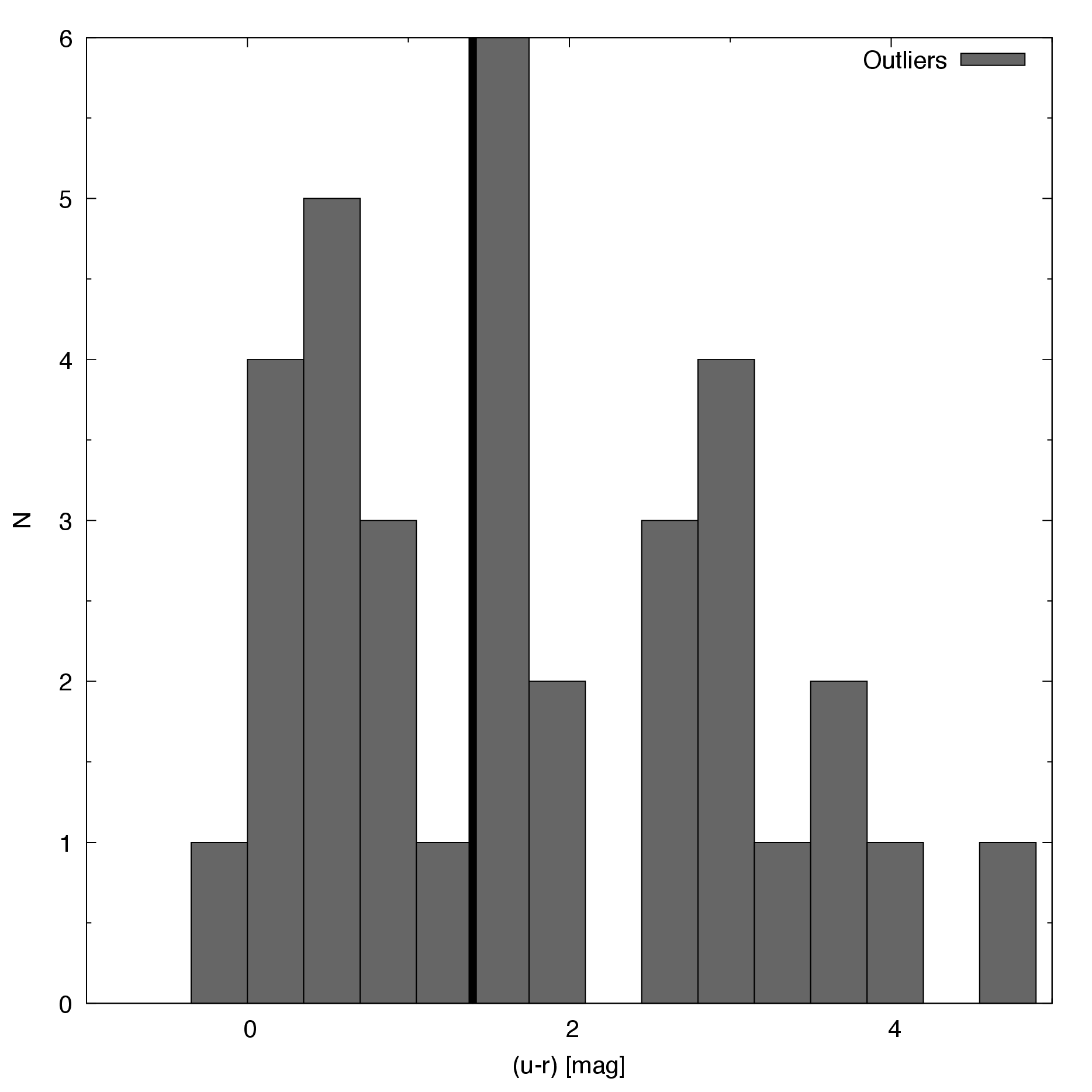}
\caption{Distribution of the (u-r) optical color taken from SDSS, for all OUTs with an optical counterpart. The vertical solid line marks the limit below which sources are expected to be of the BL Lac type, following \citet{Massaro12}.}
\end{figure}

When crossmatching the OUTs sample with sources identified or associated in our campaigns or in the literature, we found a classification for only five of them: two BZQs, two QSOs and one BZB. We show them plotted in Figure 19; they all  indeed lie closer to the BZB region of the plot than the majority of the OUTs. We note that there are 13 OUTs lying within the BZB area and above the 68\% $\rm{F}_{\rm{X}}$ threshold reported in Figure 14. None of these have been associated through optical spectroscopy, and are likely BZB candidates for future optical follow-up campaigns.

As for XBCs, we selected a subsample of OUTs that are likely BZBs based on the same criteria stated in Sect. 4.4.1 and discarding any sources that had already been associated. We obtained a subsample of 11 OUTs, which we listed in Table 4.

\begin{figure}[ht!]
\centering
\includegraphics[scale=0.45,angle=270]{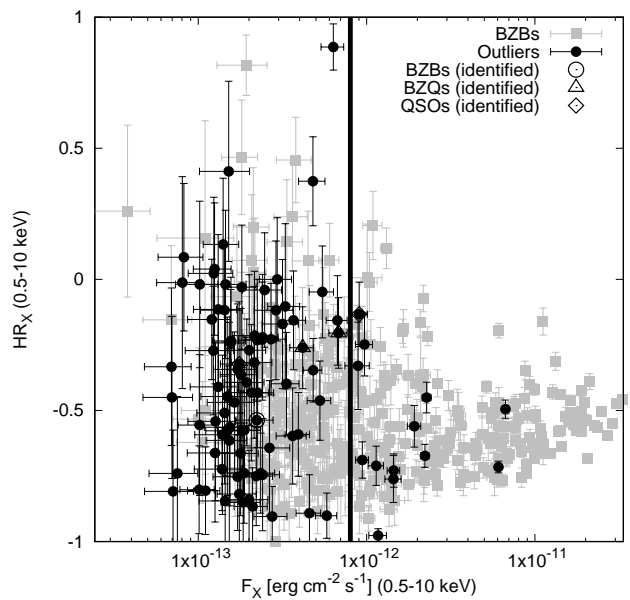}
\caption{Same as Figure 10, but including the associated sources.}
\end{figure}

\begin{table*}
\centering
\caption{Refined sample of XBCs and outliers that are with high confidence likely BL Lac sources. In column 1 we report the 3FGL name, in column 2 the \emph{Swift}/XRT source designation, in column 3 the angular separation between the 3FGL and XRT positions, and in column 4 the source class.}
\begin{tabular}{|l|l|c|c|}
\hline
Name 3FGL &  Name XRT & Ang. Sep. & Class \\
\hline\hline
  3FGLJ0031.6+0938 & SWXRTJ003159.7+093617 & 5.20 & XBC\\
  3FGLJ0430.1-3103 & SWXRTJ042958.8-305935 & 4.99 & XBC\\
  3FGLJ0516.6+1012 & SWXRTJ051641.5+101244 & 0.39 & XBC\\
  3FGLJ0802.3-0941 & SWXRTJ080216.0-094210 & 1.44 & XBC\\
  3FGLJ1117.7+0217 & SWXRTJ111731.0+021555 & 3.76 & XBC\\
  3FGLJ1119.8-2647 & SWXRTJ111957.1-264320 & 4.20 & XBC\\
  3FGLJ1120.6+0713 & SWXRTJ112042.5+071313 & 0.35 & XBC\\
  3FGLJ1200.9-1432 & SWXRTJ120055.0-143041 & 1.85 & XBC\\
  3FGLJ1225.4-3448 & SWXRTJ122536.8-344722 & 2.54 & XBC\\
  3FGLJ1228.4-0317 & SWXRTJ122819.5-031813 & 1.87 & XBC\\
  3FGLJ1249.5-0546 & SWXRTJ124919.6-054539 & 3.17 & XBC\\
  3FGLJ1309.0+0347 & SWXRTJ130832.3+034410 & 8.07 & XBC\\
  3FGLJ1513.3-3719 & SWXRTJ151318.7-372009 & 1.03 & XBC\\
  3FGLJ1810.7+5335 & SWXRTJ181037.9+533502 & 0.95 & XBC\\
  3FGLJ1934.2+6002 & SWXRTJ193419.8+600141 & 1.20 & XBC\\
  3FGLJ2043.6+0001 & SWXRTJ204342.3+000119 & 1.53 & XBC\\
  3FGLJ2044.0+1035 & SWXRTJ204351.6+103406 & 2.46 & XBC\\
  3FGLJ2047.9-3119 & SWXRTJ204806.1-312015 & 2.29 & XBC\\
  3FGLJ2142.7+1957 & SWXRTJ214247.4+195812 & 1.16 & XBC\\
 \hline
  3FGLJ0216.0+0300 & SWXRTJ021541.3+030431 & 6.51 & OUT\\
  3FGLJ0536.4-3347 & SWXRTJ053629.2-334302 & 4.37 & OUT\\
  3FGLJ0609.7-1841 & SWXRTJ061000.4-183754 & 5.67 & OUT\\
  3FGLJ0737.8-8245 & SWXRTJ073706.5-824841 & 3.33 & OUT\\
  3FGLJ0919.4+6604 & SWXRTJ091913.0+655639 & 8.03 & OUT\\
  3FGLJ0921.6+2339 & SWXRTJ092145.4+233550 & 3.74 & OUT\\
  3FGLJ0941.0+6151 & SWXRTJ094151.4+615100 & 5.45 & OUT\\
  3FGLJ1200.4+0202 & SWXRTJ120056.1+020803 & 9.23 & OUT\\
  3FGLJ1611.9+1404 & SWXRTJ161137.0+141051 & 7.56 & OUT\\
  3FGLJ1809.0+3517 & SWXRTJ180914.8+350903 & 8.29 & OUT\\
  3FGLJ2112.5-3044 & SWXRTJ211211.8-304541 & 5.18 & OUT\\
\hline\hline\end{tabular}
\end{table*}

\section{Summary and Conclusions}
\label{sec:conclusions}

In this work we analyze the X-ray properties of \,\emph{Fermi} UGSs, in particular to search for BZB candidates through the recently proposed X-ray--$\gamma$-ray connection, and comparing it with the already known connections for BZBs between $\gamma$-rays and radio, infrared, and optical wavelengths. To this aim, we built a sample of 327 UGSs listed by \,\emph{Fermi} that were observed by the \emph{Swift}/XRT telescope, collected up to December 2018. 

There are 223 out of 327 \emph{Fermi} UGSs observed by \emph{Swift}/XRT that present at least one X-ray detection in the XRT field. Out of these 223 UGSs, 148 UGSs have at least one XRT source within their positional uncertainty ellipse: We call these \emph{X-Ray blazar candidates}, or XBCs. They amount to 175 sources. A further 69 UGSs do not have any X-ray source within their positional uncertainty ellipse, but have at least one X-ray source in the field. This second sample, labeled as OUTs, lists 105 X-ray sources.

XBCs present X-ray fluxes, hardness ratio values, and mid-infrared and optical colors similar to BZBs, strongly suggesting a BZB nature. OUTs show a combination of BZB and nonBZB characteristics, indicating there are a number of BZBs within the sample but with a high degree of contamination.

There are 45 XBCs which were already classified as BZBs through optical spectroscopy. Of these, 12 ($\sim$27\%) do not show canonical mid-infrared or radio properties. This indicates that the selection of XBCs is a strong method to find $\gamma$-ray BZB candidates, that are not found with other multiwavelength
methods, within \emph{Fermi} UGSs.

Based on a combination of their X-ray and multiwavelength properties, we selected a sample of X-ray sources that are very likely BL Lac candidates. These constitute a list of 32 sources that merit follow-up optical spectroscopic observations to confirm their nature.

%  \item In particular, 69\% of the XBCs show $\rm{F}_{\gamma}>3\times10^{-12}\,\rm{erg}\,\rm{cm}^{-2}\rm{s}^{-1}$. We expect to detect 96\% of all BZBs emitting above this threshold.
%  \item 50 of the XBCs have already been classified in the literature through optical spectroscopy, 45 as BZBs, 3 as QSOs and 2 as BZQs. \item 68\% of all XBCs with X-ray fluxes $\rm{F}_{\rm{X}}\geq8\times10^{-13}\,\rm{erg}\,\rm{cm}^{-2}\rm{s}^{-1}$ were confirmed to be BZB through optical spectroscopy.
%  \item 
%  \item 
%  \item Most OUTs that follow the X-ray and mid-infrared BZB trends have a radio counterpart.
%  \item $\sim$20\% of the OUTs have X-ray hardness ratio %values and mid-infrared colors compatible with BZBs. In particular, 13 %OUTs show $\rm{F}_{\rm{X}}$ values above the 68\% threshold %established for XBCs, these are likely BZB candidates. Optical %follow-up observations are needed to confirm our predictions.
%

%
%%% This command is needed to show the entire author+affilation list when
%%% the collaboration and author truncation commands are used.  It has to
%%% go at the end of the manuscript.
%%\allauthors
%
%%% Include this line if you are using the \added, \replaced, \deleted
%%% commands to see a summary list of all changes at the end of the article.
%%\listofchanges

\section{Acknowledgements}
E. J. Marchesini would like to thank the anonymous referee on behalf of all the authors for the insights given to improve this manuscript. E. J. Marchesini would also like to thank Dr. Roc\'io I. P\'aez and Dr. M. Victoria Reynaldi for useful discussions on this work. This work is supported by the ``Departments of Excellence 2018 - 2022'' Grant awarded by the Italian Ministry of Education, University and Research (MIUR) (L. 232/2016). This research has made use of resources provided by the Compagnia di San Paolo for the grant awarded on the BLENV project (S1618\_L1\_MASF\_01) and by the Ministry of Education, Universities and Research for the grant MASF\_FFABR\_17\_01. F.M. acknowledges financial contribution from the agreement ASI-INAF n.2017-14-H.0 A.P. acknowledges financial support from the Consorzio Interuniversitario per la fisica Spaziale (CIFS) under the agreement related to the grant MASF\_CONTR\_FIN\_18\_02. This research has made use of data obtained from the high- energy Astrophysics Science Archive Research Center (HEASARC) provided by NASA's Goddard Space Flight Center. Part of this work is based on the NVSS (NRAO VLA Sky Survey). The National Radio Astronomy Observatory is operated by Associated Universities, Inc., under contract with the National Science Foundation. The Molonglo Observatory site manager, Duncan Campbell- Wilson, and the staff, Jeff Webb, Michael White, and John Barry, are responsible for the smooth operation of the Molonglo Observatory Synthesis Telescope (MOST) and the day-to-day observing program of SUMSS. SUMSS is dedicated to Michael Large, whose expertise and vision made the project possible. The MOST is operated by the School of Physics with the support of the Australian Research Council and the Science Foundation for Physics within the University of Sydney. This publication makes use of data products from the Wide-field Infrared Survey Explorer, which is a joint project of the University of California, Los Angeles, and the Jet Propulsion Laboratory/California Institute of Technology, funded by the National Aeronautics and Space Administration. TOPCAT\footnote{http://www.star.bris.ac.uk/m~bt/topcat/} \citep{Taylor05} and STILTS \citep{Taylor06} were used for the preparation and manipulation of the images and the tabular data.

%\newpage
\bibliography{Biblio} 
\bibliographystyle{aa}
%\newpage

\end{document}